# The Nu Class of Low-Degree-Truncated, Rational, Generalized Functions. Ib. Integrals of Matérn-correlation functions for all odd–half-integer class parameters

Selden Crary[1], Richard Diehl Martinez[2], and Michael Saunders[2]

*1. Palo Alto, CA, USA*

*2. Department of Management Science and Engineering*
*Stanford University, Stanford, CA, USA*

**Abstract**

This paper is an extension of Parts I and Ia of a series about Nu-class generalized functions. We provide hand-generated algebraic expressions for integrals of single Matérn-covariance functions, as well as for products of two Matérn-covariance functions, for all odd-half-integer class parameters. These are useful both for IMSPE-optimal design software and for testing universality of Nu-class, generalized-function properties, across covariance classes.

## 1. Introduction

The present paper is Part Ib of the Roman-numeralled series Parts I through V, with new-Latin–lowercase-lettered sub-parts, reporting research into Nu-class generalized functions [1].

We begin with a short history of relevant earlier papers in this series.

Part I, Appendices H and J [1] included lengthy, verbatim, closed-form, symbolic-manipulation-software–generated, algebraic expressions of one-dimensional integrals, named $I_5$ and $I_7$, of single Matérn-covariance functions, with respective class parameters $\nu = 3/2$ and $5/2$. In the same paper, Appendices I and K included integrals, named $I_6$ and $I_8$, for the respective products of two identical-class Matérn-covariance functions. These last two integrals were left in rough, symbolic-manipulation-software output format. These expressions removed the need for computationally intensive Monte-Carlo approximations that previously had been common practice [3], when the integrals were required, as in the computation of an IMSPE-optimal design based on one of these Matérn-covariance functions.

Part Ia [2] then provided succinct algebraic expressions for these two integrals, more suitable, compared to those in Paper I, for software development involving computation of the IMSPE. However, the expressions were restricted to class parameters $3/2$ and $5/2$, and their correctness depended upon the correctness of the outputs of the original symbolic-manipulation software, thus ruling out rigorous mathematical proofs based on these expressions.

The present Paper Ib provides exclusively–hand-generated algebraic expressions of products of Matérn-covariance functions, for all odd-half-integer class parameters. These expressions agree with those given in Parts I and Ia. The general expressions are algebraic sums succinct enough for use in software development for moderate-size class parameters.

In a subsequent part, the expressions derived here shall be used to test mathematically rigorous universality of Nu-cla,ss generalized-function properties, across covariance classes. For example, we shall test for the universal occurrence of quantum phase transitions of designs for computer experiments, across their hyper-parameter spaces [4].

## 2. Outline

## 3. List of tables

# 4. List of example integrals $I_{\nu=p+1/2}$ and $J_{\nu=p+1/2}$

# 5. Identities and elementary integrals

## 5.1 Double-factorial identity

$$\frac{(2p)!}{p!} = (2p-1)!!\,2^p,\ \ p \in \mathbb{N}_{\geq 0}, \tag{5.1}$$

where $n!! \equiv \prod_{i=0}^{\lceil n/2 \rceil - 1}(n-2i) = n(n-2)(n-4)\cdots$, and $0!! \equiv (-1)!! \equiv 1$.

Demonstration:

$$\frac{(2p)!}{p!} = \frac{(2p)(2p-1)\cdots 1}{p!} = \frac{\overbrace{[(2p)(2p-2)\cdots 2]}^{p\ \text{even factors}}\cdot[(2p-1)(2p-3)\cdots 1]}{p!} = \frac{2^p p!\cdot[(2p-1)(2p-3)\cdots 1]}{p!} = 2^p(2p-1)!!.$$

## 5.2 Variation of Finite Companion Binomial theorem

$$\sum_{j=0}^{p} \frac{(2p-j)!}{(p-j)!} 2^j = 4^p p!\,,\ \ p \in \mathbb{N}_{\geq 0}. \tag{5.2}$$

Demonstration: This follows from the "Variation of Finite Companion Binomial Theorem" tabulated, albeit unproved, in [5] as $\sum_{k=0}^{n} \binom{2n-k}{n} 2^k = 2^{2n}$, where $\binom{2n-k}{n} \equiv \frac{(2n-k)!}{n!(n-k)!}$. Changing variables $[n,k]$ to $[p,j]$ and identifying $2^{2p} = 4^p$ gives Eq. 5.2. The Appendix of the present paper provides a detailed graphical demonstration.

## 5.3 Fubini-principle identity

$$\sum_{j=0}^{p} f(j) \sum_{m=0}^{j} g(m) = \sum_{m=0}^{p} g(m) \sum_{j=m}^{p} f(j)\ \ [6]. \tag{5.3}$$

Example: $\sum_{j=0}^{p} \frac{(2p-j)!}{(p-j)!} 2^j \sum_{m=0}^{j} \frac{\sqrt{(2p+1)\theta}^{\,m}(1+a)^m}{m!} = \sum_{m=0}^{p} \frac{\sqrt{(2p+1)\theta}^{\,m}(1+a)^m}{m!} \sum_{j=m}^{p} \frac{(2p-j)!}{(p-j)!} 2^j$.

## 5.4 Symmetry operators

We reintroduce the algebraic symmetry operators $\mathcal{S}_w$ and $\mathcal{T}_{a,b}^{(+)}$ that were defined in Part I, Sub-appendix R.5 [1] and Part Ia, Section 3 [2], respectively:

$\mathcal{S}_w f(w,x) \equiv f(-w,x)$ changes the sign on all algebraic quantities denoted by $w$, e.g., $\mathcal{S}_w(a + bw + cw^2 + dx + ex^2) = a - bw + cw^2 + dx + ex^2$, where $x$ has no dependence on $w$.

$\mathcal{T}_{a;b}^{(+)} \equiv \mathcal{I} + \mathcal{S}_a\mathcal{S}_b$, where $\mathcal{I}$ is the identity operator. Example: $\mathcal{T}_{a;b}\left(1 + \frac{a+b}{2}\right) = 2$.

We also define additional algebraic symmetry operators, $\mathcal{T}_a^{(+)}$ and $\mathcal{T}_{a;b}^{(-)}$, as follows:

$\mathcal{T}_a^{(+)} \equiv \mathcal{I} + \mathcal{S}_a$. $\quad \mathcal{T}_{x_{i,k};x_{j,k}}$ is in Eqs. 9.1 & 9. (5.4)

$\mathcal{T}_{a;b}^{(-)} \equiv \mathcal{I} - \mathcal{S}_a\mathcal{S}_b$. (5.5)

### 5.5 Elementary integrals

Dwight 567.9 gives almost literally [7]

$$\int x^n e^{bx} dx = e^{bx}\left[\frac{x^n}{b} - \frac{nx^{n-1}}{b^2} + \frac{n(n-1)x^{n-2}}{b^3} - \cdots + (-1)^{n-1}\frac{n!x}{b^n} + (-1)^n\frac{n!}{b^{n+1}}\right], \quad n \geq 0.$$

Specific examples are the following:

*for* $[n,b] = [0,-1]$:

$$\int e^{-x} dx = -e^{-x};$$

*for* $[n,b] = [j \geq 1, -1]$:

$$\int x^j e^{-x} dx = -e^{-x}\left[j! + j!\,x + \cdots + j(j-1)x^{j-2} + jx^{j-1} + x^j\right]$$

$$= -e^{-x}\,j!\left[\frac{x^0}{0!} + \frac{x^1}{1!} + \cdots + \frac{x^{j-2}}{(j-2)!} + \frac{x^{j-1}}{(j-1)!} + \frac{x^j}{j!}\right]$$

$$= -e^{-x} j! \sum_{m=0}^{j} \frac{x^m}{m!}; \text{ and} \tag{5.6}$$

*for* $[n,b] = [j+k-l, -2]$ (These specific values will be used between Eqs. 8.6 and 8.7.)

$$\int x^{j+k-l}\, e^{-2x} dx = e^{-2x}\left[\begin{array}{l} \frac{x^{j+k-l}}{b} - \frac{(j+k-l)x^{j+k-l-1}}{b^2} + \frac{(j+k-l)(j+k-l-1)x^{j+k-l-2}}{b^3} - \cdots \\ \qquad + (-1)^{j+k-l-1}\frac{(j+k-l)!x}{b^{j+k-l}} + (-1)^{j+k-l}\frac{(j+k-l)!}{b^{j+k-l+1}} \end{array}\right]$$

$$= e^{-2x}(j+k-l)!\left[\begin{array}{l} \frac{x^{j+k-l}}{(j+k-l)!b} - \frac{(j+k-l)x^{j+k-l-1}}{(j+k-l)!b^2} + \frac{(j+k-l)(j+k-l-1)x^{j+k-l-2}}{(j+k-l)!b^3} - \cdots \\ \qquad + (-1)^{j+k-l-1}\frac{x}{b^{j+k-l}1!} + (-1)^{j+k-l}\frac{1}{b^{j+k-l+1}0!} \end{array}\right]$$

$$= e^{-2x}(j+k-l)!\left[\begin{array}{l} \frac{x^{j+k-l}}{(j+k-l)!b} - \frac{x^{j+k-l-1}}{(j+k-l-1)!b^2} + \frac{x^{j+k-l-2}}{(j+k-l-2)!b^3} - \cdots \\ \qquad + (-1)^{j+k-l-1}\frac{x^1}{b^{j+k-l}} + (-1)^{j+k-l}\frac{x^0}{b^{j+k-l+1}} \end{array}\right]$$

$$= e^{-2x}(j+k-l)! \sum_{m=0}^{j+k-l} \frac{(-1)^{j+k-l-\{}x^m}{m!b^{j+k-l-\{+1}}$$

$$= -\frac{(j+k-l)!}{2^{j+k-l+1}} e^{-2x} \sum_{m=0}^{j+k-l} \frac{(2x)^m}{m!}. \tag{5.7}$$

## 6. Integrals of single Matérn-correlation functions

Our interest is the integration of odd-half-integer-parameter Matérn-correlation functions defined as functions of a radial distance $r$ as the following, which is taken literally from Eq. 4.16 of Rasmussen and Williams [8]:

$$K_{\nu=p+1/2}(r) = exp\left(\frac{-\sqrt{2\nu}r}{\ell}\right)\frac{\Gamma(p+1)}{\Gamma(2p+1)}\sum_{i=0}^{p}\frac{(p+i)!}{i!(p-i)!}\left(\frac{\sqrt{8\nu}r}{\ell}\right)^{p-i}, \tag{6.1}$$

where $p \in \mathbb{N}_{\geq 0}$, $\nu$ is as defined on the left-hand side of Eq. 6.1, and $\ell$ is a positive parameter.

For the purposes of the present section, we are interested in this correlation as a function of the distance between an arbitrary Cartesian-coordinate location $x$ and the $k$'th Cartesian coordinate of the $i$'th design point, i.e., $x_{i,k}$. Throughout most of this section, we use the following definitions:

$$a \equiv x_{i,k} \text{ and} \tag{6.2}$$

$$\theta \equiv \theta_k. \tag{6.3}$$

After using the property of gamma functions of natural numbers $\Gamma(p+1) = p!$ and making connection with the covariance hyper-parameter used in Parts I and Ia of this series of papers [1,2], via $\theta \equiv 1/\ell^2$, Eq. 6.1 becomes

$$K_{p+1/2}(|a-x|) = \frac{p!}{(2p)!}\sum_{i=0}^{p}\frac{(p+i)!}{i!(p-i)!}\left(\sqrt{4(2p+1)\theta|a-x|^2}\right)^{p-i} e^{-\sqrt{(2p+1)\theta|a-x|^2}}.$$

Defining $j \equiv p - i$, reversing the order in the summation, and rearranging slightly, gives

$$K_{p+1/2}(|a-x|) = \frac{p!}{(2p)!}\sum_{j=0}^{p}\frac{(2p-j)!}{(p-j)!j!}2^j\sqrt{(2p+1)\theta|a-x|^2}^{\,j} e^{-\sqrt{(2p+1)\theta|a-x|^2}}.$$

The two most frequently discussed such functions in the statistics literature [9,10] are the following:

for $p = 1$: $K_{3/2}(|a-x|) \equiv \left(1+\sqrt{3\theta|a-x|^2}\right)e^{-\sqrt{3\theta|a-x|^2}}$, and

for $p = 2$: $K_{5/2}(|a-x|) \equiv \left(1+\sqrt{5\theta|a-x|^2}+\frac{5\theta|a-x|^2}{3}\right)e^{-\sqrt{5\theta|a-x|^2}}$.

The definite integrals of these functions over the range $-1 \leq x \leq 1$ are the subject of interest:

$$I_{p+1/2}(a,\theta) = \frac{p!}{(2p)!}\sum_{j=0}^{p}\frac{(2p-j)!}{(p-j)!j!}2^j\frac{1}{2}\left(\begin{array}{c}\int_{-1}^{a}\sqrt{(2p+1)\theta|a-x|^2}^{\,j}e^{-\sqrt{(2p+1)\theta|a-x|^2}}dx \\ +\int_{a}^{1}\sqrt{(2p+1)\theta|a-x|^2}^{\,j}e^{-\sqrt{(2p+1)\theta|a-x|^2}}dx\end{array}\right). \tag{6.4}$$

We introduce the following changes in variables:

for $x < a$: $\tilde{x} \equiv [(2p+1)\theta]^{1/2}(a-x)$, $\quad d\tilde{x} = -[(2p+1)\theta]^{1/2}dx$, and

$$[(2p+1)\theta|a-X|^2]^{j/2} = [(2p+1)\theta]^{j/2}(a-x)^j = \tilde{x}^j;$$

for $x \geq a$: $\tilde{\tilde{x}} \equiv [(2p+1)\theta]^{1/2}(x-a)$, $\quad d\tilde{\tilde{x}} = [(2p+1)\theta]^{1/2}dx$, and

$$[(2p+1)\theta|a-X|^2]^{j/2} = [(2p+1)\theta]^{j/2}(x-a)^j = \tilde{\tilde{x}}^j.$$

Reversing the limits on the second integral in Eq. 6.4 gives

$$I_{p+1/2}(a,\theta) = -\frac{p!}{2\sqrt{(2p+1)\theta}(2p)!}\sum_{j=0}^{p}\frac{(2p-j)!}{(p-j)!j!}2^j\begin{pmatrix} \int_{\tilde{x}=\sqrt{(2p+1)\theta}\,(1+a)}^{0} \tilde{x}^j e^{-\tilde{x}}\,d\tilde{x} \\ +\int_{\tilde{\tilde{x}}=\sqrt{(2p+1)\theta}\,(1-a)}^{0} \tilde{\tilde{x}}^j e^{-\tilde{\tilde{x}}}\,d\tilde{\tilde{x}} \end{pmatrix}.$$

Using the twin operator $\mathcal{T}_a^{(+)}$ of Eq. 5.4 and dropping all tildes gives

$$I_{p+1/2}(a,\theta) = -\frac{p!}{2\sqrt{(2p+1)\theta}(2p)!}\sum_{j=0}^{p}\frac{(2p-j)!}{(p-j)!j!}2^j\mathcal{T}_a^{(+)}\int_{x=\sqrt{(2p+1)\theta}(1+a)}^{0} x^j e^{-x}\,dx. \tag{6.5}$$

Substituting the elementary integral of $x^j e^{-x}$ of Eq. 5.6 into Eq. 6.5, canceling $j!$'s, and moving left both $\mathcal{T}_a^{(+)}$ and $e^{-x}$ gives

$$I_{p+1/2}(a,\theta) = \frac{p!}{2\sqrt{(2p+1)\theta}(2p)!}\mathcal{T}_a^{(+)}e^{-x}\sum_{j=0}^{p}\frac{(2p-j)!}{(p-j)!}2^j\sum_{m=0}^{j}\frac{x^m}{m!}\Big|_{x=\sqrt{(2p+1)\theta}(1+a)}^{0}.$$

Noting the upper evaluation value $x = 0$ contributes only via $0^0 = 1$ gives

$$I_{p+1/2}(a,\theta) = \frac{p!}{2\sqrt{(2p+1)\theta}(2p)!}\mathcal{T}_a^{(+)}\begin{bmatrix} \sum_{j=0}^{p}\frac{(2p-j)!}{(p-j)!}2^j \\ -e^{-\sqrt{(2p+1)\theta}(1+a)}\sum_{j=0}^{p}\frac{(2p-j)!}{(p-j)!}2^j\sum_{m=0}^{j}\frac{\sqrt{(2p+1)\theta}^{\,m}(1+a)^m}{m!} \end{bmatrix}.$$

Reversing the order of the double sum via the Fubini-principle identity of Eq. 5.3, gives

$$I_{p+1/2}(a,\theta) = \frac{p!}{2\sqrt{(2p+1)\theta}(2p)!}\mathcal{T}_a^{(+)}\begin{bmatrix} \sum_{j=0}^{p}\frac{(2p-j)!}{(p-j)!}2^j \\ -\sum_{m=0}^{p}\frac{\sqrt{(2p+1)\theta}^{\,m}(1+a)^m}{m!}\sum_{j=m}^{p}\frac{(2p-j)!}{(p-j)!}2^j\,e^{-\sqrt{(2p+1)\theta}(1+a)} \end{bmatrix}.$$

Expanding the second sum and dropping the explicit argument of $I_{p+1/2}$ gives

$$I_{p+1/2} = \frac{p!}{2\sqrt{(2p+1)\theta}(2p)!}\mathcal{T}_a^{(+)}\left(\begin{array}{c} \sum_{j=0}^{p}\frac{(2p-j)!}{(p-j)!}2^j \\ -\left\{\begin{array}{c} \frac{1}{0!}\sum_{j=0}^{p}\frac{(2p-j)!}{(p-j)!}2^j\sqrt{(2p+1)\theta}^{\,0}(1+a)^0 \\ +\frac{1}{1!}\left[\sum_{j=1}^{p}\frac{(2p-j)!}{(p-j)!}2^j\right]\sqrt{(2p+1)\theta}^{\,1}(1+a)^1 \\ +\frac{1}{2!}\left[\sum_{j=2}^{p}\frac{(2p-j)!}{(p-j)!}2^j\right]\sqrt{(2p+1)\theta}^{\,2}(1+a)^2 \\ \vdots \\ +2^p\sqrt{(2p+1)\theta}^{\,p}(1+a)^p \end{array}\right\}e^{-\sqrt{(2p+1)\theta}(1+a)} \end{array}\right).$$

Using the double-factorial identity $\frac{p!}{(2p)!} = \frac{1}{(2p-1)!!2^p}$ from Eq. 5.1, using subscripts on square brackets to denote evaluation at specific values of $j$, and replacing $\theta$ and $a$ with $\theta_k$ and $x_{i,k}$, respectively, gives

$$
I_{p+1/2} = \frac{1}{2(2p-1)!!\sqrt{(2p+1)\theta_k}}\mathcal{T}^{(+)}_{x_{i,k}}\left[\begin{array}{c}
\frac{1}{2^p}\sum_{j=0}^{p}\frac{(2p-j)!}{(p-j)!}2^j \\
-\left(\begin{array}{c}
\frac{1}{0!2^p}\left[\sum_{j=0}^{p}\frac{(2p-j)!}{(p-j)!}2^j\right]\sqrt{(2p+1)\theta_k}^{\,0}\left(1+x_{i,k}\right)^0 \\
+\frac{1}{1!2^p}\left\{\begin{array}{c}\sum_{j=0}^{p}\frac{(2p-j)!}{(p-j)!}2^j \\ -\left[\frac{(2p-j)!}{(p-j)!}2^j\right]_{j=0}\end{array}\right\}\sqrt{(2p+1)\theta_k}^{\,1}\left(1+x_{i,k}\right)^1 \\
+\frac{1}{2!2^p}\left\{\begin{array}{c}\sum_{j=0}^{p}\frac{(2p-j)!}{(p-j)!}2^j \\ -\left[\frac{(2p-j)!}{(p-j)!}2^j\right]_{j=0} \\ -\left[\frac{(2p-j)!}{(p-j)!}2^j\right]_{j=1}\end{array}\right\}\sqrt{(2p+1)\theta_k}^{\,2}\left(1+x_{i,k}\right)^2 \\
\vdots \\
+\sqrt{(2p+1)\theta_k}^{\,p}\left(1+x_{i,k}\right)^p
\end{array}\right)e^{-\sqrt{(2p+1)\theta_k}\left(1+x_{i,k}\right)}
\end{array}\right].
$$

The sums with index $j$ can be removed by Identity 5.2, $\sum_{j=0}^{p}\frac{(2p-j)!}{(p-j)!}2^j = 4^p p!$. Then, evaluating the ultimate equation for its stated values of $j$ gives

$$
I_{p+1/2} = \frac{1}{2(2p-1)!!\sqrt{(2p+1)\theta_k}}\mathcal{T}^{(+)}_{x_{i,k}}\left(\begin{array}{c}
\frac{1}{2^p}(4^p p!) \\
-\left\{\begin{array}{c}
\frac{1}{0!2^p}(4^p p!)\quad [(2p+1)\theta_k]^{0/2}\left(1+x_{i,k}\right)^0 \\
+\frac{1}{1!2^p}\left[\begin{array}{c}4^p p! \\ -\frac{(2p)!}{(p)!}\end{array}\right][(2p+1)\theta_k]^{1/2}\left(1+x_{i,k}\right)^1 \\
+\frac{1}{2!2^p}\left[\begin{array}{c}4^p p! \\ -\frac{(2p)!}{(p)!} \\ -\frac{(2p-1)!}{(p-1)!}2\end{array}\right][(2p+1)\theta_k]^{2/2}\left(1+x_{i,k}\right)^2 \\
\vdots \\
+[(2p+1)\theta_k]^{p/2}\left(1+x_{i,k}\right)^p
\end{array}\right\}e^{-\sqrt{(2p+1)\theta_k}\left(1+x_{i,k}\right)}
\end{array}\right).
$$

Expressing this with the natural-number functions,

$a_0(p) \equiv \frac{1}{2^p}4^p p! = 2^p p!$ and

$b_j(p) \equiv \frac{1}{2^p}\frac{(2p-j)!}{(p-j)!}2^j,\ \ 0 \le j \le p-1$, gives

$$
I_{p+1/2} = \frac{1}{2(2p-1)!!\sqrt{(2p+1)\theta_k}}\mathcal{T}^{(+)}_{x_{i,k}}\left\{\begin{array}{c}
a_0 \\
-\left[\begin{array}{c}
a_0\sqrt{(2p+1)\theta_k}^{\,0}\left(1+x_{i,k}\right)^0 \\
+\frac{1}{1!}(a_0-b_0)\sqrt{(2p+1)\theta_k}^{\,1}\left(1+x_{i,k}\right)^1 \\
+\frac{1}{2!}(a_0-b_0-b_1)\sqrt{(2p+1)\theta_k}^{\,2}\left(1+x_{i,k}\right)^2 \\
\vdots \\
+\sqrt{(2p+1)\theta_k}^{\,p}\left(1+x_{i,k}\right)^p
\end{array}\right]e^{-\sqrt{(2p+1)\theta_k}\left(1+x_{i,k}\right)}
\end{array}\right\}. \tag{6.6}
$$

The natural numbers can be computed simply and recursively from Table 1, immediately below, or picked off directly from Table 2.

N.B.: Identity 5.2, viz., $\sum_{j=0}^{p} \frac{(2p-j)!}{(p-j)!} 2^j = 4^p p!$, was used to simplify the left heading in Table 1 to $2^p p!$

| | $a_0(p) \equiv 2^p p!$ | $b_j(p) \equiv \frac{1}{2^p} \frac{(2p-j)!}{(p-j)!} 2^j$ | | | | | | |
|---|---|---|---|---|---|---|---|---|
| | | *j* | | | | | | |
| *p* | | *0* | *1* | *2* | *3* | *4* | *5* | *6* |
| *1* | 2 | 1 | 1 | | | | | |
| *2* | 8 | 3 | 3 | 2 | | | | |
| *3* | 48 | 15 | 15 | 12 | 6 | | | |
| *4* | 384 | 105 | 105 | 90 | 60 | 24 | | |
| *5* | 3840 | 945 | 945 | 840 | 630 | 360 | 120 | |
| *6* | 46080 | 10395 | 10395 | 9450 | 7560 | 5040 | 2520 | 720 |
| ⋮ | ⋮ | ⋮ | ⋮ | ⋮ | ⋮ | ⋮ | ⋮ | ⋮ |
| *12 wrapped* | 1 961 990 553 600 | 316 234 143 225 | 316 234 143 225 | 302 484 832 650 | 274 986 211 500 | 235 702 467 000 | 188 561 973 600 | 138 940 401 600 |
| *OEIS* | A002866 | A001147 | A001147 | A001879 | A000457 | A001880 | A001881 | A038121 |

Table 1. Tabulated values of the natural numbers $a_0$ and $b_j$ appearing in Eq. 6.6, as functions of $p$. The putatively identical sequence found in the OEIS [11] is given for each column. Along the right-most diagonal, i.e., for $j = p$, $b_p(p) = p! = p \cdot b_{p-1}(p-1)$. Along the k'th-penultimate-RHS main diagonal, $b_{p-k}(p)! = (p+k) \cdot b_{p-k-1}(p-1)$.

| | | | $\frac{1}{(k+1)!}\left(a_0 - \sum_{j=0}^{k} b_j\right)$ | | | | | |
|---|---|---|---|---|---|---|---|---|
| | | | $k$ | | | | | |
| *p* | $a_0$ | $a_0$ | *0* | *1* | *2* | *3* | *4* | *5* |
| *1* | 2 | 2 | 1 | | | | | |
| *2* | 8 | 8 | 5 | 1 | | | | |
| *3* | 48 | 48 | 33 | 9 | 1 | | | |
| *4* | 384 | 384 | 279 | 87 | 14 | 1 | | |
| *5* | 3840 | 3840 | 2895 | 975 | 185 | 20 | 1 | |
| *6* | 46080 | 46080 | 35685 | 12645 | 2640 | 345 | 27 | 1 |
| ⋮ | ⋮ | ⋮ | ⋮ | ⋮ | ⋮ | ⋮ | ⋮ | ⋮ |
| *12 wrapped* | 1 961 990 553 600 | 1 961 990 553 600 | 1 645 756 410 375 | 664 761 133 575 | 171 172 905 750 | 31 335 467 625 | 4 302 906 300 | 455 259 420 |
| *OEIS* | A002866 | A002866 | A129890 | A035101 | A263384 | none | none | none |

Table 2. Tabulated values of the sequential natural numbers $a_0$ (twice), $\frac{1}{1!}(a_0 - b_0)$, $\frac{1}{2!}(a_0 - b_0 - b_1)$, $\cdots$ appearing in Eq. 6.6, as calculated from Table 1 and as functions of $p$. The putatively corresponding sequence found in the OEIS [11] is given for most columns. The sequences of numbers in the rows do not lead to OEIS entries. For example, the reversed $p = 6$ row, viz., (1, 27, 345, 2640, 12645, 35685, 46080), is not in the OEIS. Details for numbers appearing on the upper-left-to-lower-right diagonals for $k \geq 0$: the rightmost diagonal elements are all unity; while the next-to-rightmost diagonal elements, viz., (5, 9, 14, 20, 27), follow the sequence $(p+1)(p+2)/2 - 1$, $p \geq 2$, which is the sequence of triangle numbers minus unity. The repeated $a_0$ column was included to assist visualization of the following two observations: the extended-to-the-left-by-one-column, next-to-rightmost diagonal elements, viz., (2 ,5, 9, 14, 20, 27), continue to follow the sequence $(p+1)(p+2)/2 - 1$, $p \geq 2$; and the extended next-to-next-to-rightmost diagonal elements, viz., (2 ,8, 33, 87, 185, 345) may follow some still-to-be-determined rule.

## 7. Four example integrals of single Matérn-correlation functions: $I_{3/2}, I_{5/2}, I_{7/2}$, and $I_{9/2}$

We now use Eq. 6.6 in conjunction with Tables 1 and 2 to generate algebraic expressions for integrals $I_{3/2}, I_{5/2}, I_{7/2}$, and $I_{9/2}$.

**Example:** ***p*=1:** $\boldsymbol{I_{\nu=p+1/2} \equiv I_{3/2} \equiv R_{0,i}^{(p=1)}}$

Using Eq. 6.5, using $a_0$ and $b_0$ from the $p = 1$ row of Table 2, and replacing $a$ and $\theta$ with $x_{i,k}$ and $\theta_k$, respectively, via Eqs. 6.2 and 6.3, gives

$$I_{3/2}(a,\theta) = \frac{1}{2\sqrt{3\theta_k}} \mathcal{T}_{x_{i,k}}^{(+)} \left\{ a_0 - \left[ a_0 + (a_0 - b_0)\sqrt{3\theta_k}\left(1 + x_{i,k}\right) \right] e^{-\sqrt{3\theta_k}(1+x_{i,k})} \right\}, \quad 1 \leq i \leq \mathrm{n}.$$

$$= \frac{1}{2\sqrt{3\theta_k}} \mathcal{T}_{x_{i,k}}^{(+)} \left\{ 2 - \left[ 2 + \sqrt{3\theta_k}\left(1 + x_{i,k}\right) \right] e^{-\sqrt{3\theta_k}(1+x_{i,k})} \right\}. \tag{7.1}$$

This agrees, as it should, with the expression given in Part I, Table 4.2 of this series of papers, viz.,

$$R_{0,i}^{(p=1)} = \frac{1}{2\sqrt{3\theta_k}} \left\{ \begin{array}{r} 2\begin{bmatrix} 1 - e^{-\sqrt{3\theta_k}\,(1+x_{i,k})} \\ +1 - e^{-\sqrt{3\theta_k}\,(1-x_{i,k})} \end{bmatrix} \\ -\sqrt{3\theta_k}\begin{bmatrix} (1+x_{i,k})e^{-\sqrt{3\theta_k}\,(1+x_{i,k})} \\ +(1-x_{i,k})e^{-\sqrt{3\theta_k}\,(1-x_{i,k})} \end{bmatrix} \end{array} \right\}.$$

**Example:** ***p*=2:** $\boldsymbol{I_{\nu=p+1/2} \equiv I_{5/2} \equiv R_{0,i}^{(p=2)}}$

In similar fashion to the example, immediately above, but using $a_0, b_0,$ and $b_1$ from the $p = 2$ row of Table 2 gives

$$I_{5/2}(x_{i,k}, \theta_k) = \frac{1}{6\sqrt{5\theta_k}} \mathcal{T}_{x_{i,k}}^{(+)} \left\{ \begin{array}{l} a_0 \\ -\begin{bmatrix} a_0 \\ + \quad (a_0 - b_0)\ \sqrt{5\theta_k}\,(1+x_{i,k}) \\ +\frac{1}{2}(a_0 - b_0 - b_1)\ \ 5\theta_k\ \ (1+x_{i,k})^2 \end{bmatrix} e^{-\sqrt{5\theta_k}(1+x_{i,k})} \end{array} \right\}$$

$$= \frac{1}{6\sqrt{5\theta_k}} \mathcal{T}_{x_{i,k}}^{(+)} \left\{ \begin{array}{l} 8 \\ -\begin{bmatrix} 8 \\ +5\sqrt{5\theta_k}\,(1+x_{i,k}) \\ +1 \cdot 5\theta_k\ \ (1+x_{i,k})^2 \end{bmatrix} e^{-\sqrt{5\theta_k}(1+x_{i,k})} \end{array} \right\}. \tag{7.2}$$

This agrees with the expression given in Part I, Table 4.2 of this series of papers [1], viz.,

$$R_{0,i}^{(p=2)} = \frac{1}{6\sqrt{5\theta_k}} \left\{ \begin{array}{r} 8\begin{bmatrix} 1 - e^{-\sqrt{5\theta_k}\,(1+x_{i,k})} \\ +1 - e^{-\sqrt{5\theta_k}\,(1-x_{i,k})} \end{bmatrix} \\ -5\sqrt{5\theta_k}\begin{bmatrix} (1+x_{i,k})e^{-\sqrt{5\theta_k}\,(1+x_{i,k})} \\ +(1-x_{i,k})e^{-\sqrt{5\theta_k}\,(1-x_{i,k})} \end{bmatrix} \\ -5\theta_k\begin{bmatrix} (1+x_{i,k})^2 e^{-\sqrt{5\theta_k}\,(1+x_{i,k})} \\ +(1-x_{i,k})^2 e^{-\sqrt{5\theta_k}\,(1-x_{i,k})} \end{bmatrix} \end{array} \right\}.$$

*Additional examples of integrals, generated from Eq. 4.5 and Table 2, are given below:*

$$R_{0,i}^{(p=3)} = I_{7/2}(x_{i,k}, \theta_k) = \frac{1}{30\sqrt{7\theta_k}} \mathcal{T}_{x_{i,k}}^{(+)} \left\{ \begin{array}{l} 48 \\ -\begin{bmatrix} 48 \\ +33\,\sqrt{7\theta_k}\ \ (1+x_{i,k}) \\ +\ 9\,\sqrt{7\theta_k}^{\,2}\,(1+x_{i,k})^2 \\ +\ 1\,\sqrt{7\theta_k}^{\,3}\,(1+x_{i,k})^3 \end{bmatrix} e^{-\sqrt{7\theta_k}(1+x_{i,k})} \end{array} \right\}. \tag{7.3}$$

$$R_{0,i}^{(p=4)} = I_{9/2}(a,\theta) = \frac{1}{210\sqrt{9\theta_k}}\mathcal{T}_{x_{i,k}}^{(+)}\left\{\begin{matrix} 384 \\ -\begin{bmatrix} 384 \\ +279\sqrt{9\theta_k}\ \left(1+x_{i,k}\right) \\ +\ 87\sqrt{9\theta_k}^{2}\left(1+x_{i,k}\right)^2 \\ +\ 14\sqrt{9\theta_k}^{3}\left(1+x_{i,k}\right)^3 \\ +\ \ 1\sqrt{9\theta_k}^{4}\left(1+x_{i,k}\right)^4 \end{bmatrix} e^{-\sqrt{9\theta}(1+x_{i,k})} \end{matrix}\right\}. \tag{7.4}$$

# 8. Integrals of products of two Matérn-correlation functions

Sec. 6 provided algebraic expressions for integrals of single Matérn-correlation functions. In the present section we derive an expression for integrals of the products of the following two Matérn-correlation functions, as a function of the distance between an arbitrary Cartesian-coordinate location $x$ and the $k$'th Cartesian coordinates of the $i$'th and $j$'th design points, i.e., $x_{i,k}$ and $x_{j,k}$, respectively:

$$K_{p+1/2}(|a-x|) = \frac{p!}{(2p)!}\sum_{j=0}^{p}\frac{(2p-j)!}{(p-j)!j!}2^j\sqrt{(2p+1)\theta\left|x_{i,k}-x\right|^2}^{\,j}\ e^{-\sqrt{(2p+1)\theta\left|x_{i,k}-x\right|^2}} \text{ and}$$

$$K_{p+1/2}(|b-x|) = \frac{p!}{(2p)!}\sum_{k=0}^{p}\frac{(2p-k)!}{(p-k)!k!}2^k\sqrt{(2p+1)\theta\left|x_{j,k}-x\right|^2}^{\,k}\ e^{-\sqrt{(2p+1)\theta\left|x_{j,k}-x\right|^2}}.$$

Throughout most of this section, we substitute the letter $a$ for $x_{i,k}$ and the letter $b$ for $x_{j,k}$, i.e.,

$a \equiv x_{i,k}$, which repeats Eq. 6.2, and

$b \equiv x_{j,k}$. (8.1)

The definite integrals of the products of these functions over the range $-1 \le x \le 1$ and for $-1 \le a \le b \le 1$ are the subject of interest, viz.,

$$J_{p+1/2}(|a-x|,|b-x|) \equiv \frac{1}{2}\int_{-1}^{1}\left[K_{p+1/2}(|a-x|)\right]\left[K_{p+1/2}(|b-x|)\right]dx$$

$$= \left[\frac{p!}{(2p)!}\right]^2\sum_{j,k=0}^{p}\frac{(2p-j)!(2p-k)!2^{j+k}}{(p-j)!j!(p-k)!k!}$$

$$\cdot\frac{1}{2}\left\{\begin{matrix} \int_{x=-1}^{a}\left[\sqrt{(2p+1)\theta}(a-x)\right]^j\left[\sqrt{(2p+1)\theta}(b-x)\right]^k e^{-\sqrt{(2p+1)\theta}[(a-x)+(b-x)]}dx \\ +\int_{x=a}^{b}\left[\sqrt{(2p+1)\theta}(x-a)\right]^j\left[\sqrt{(2p+1)\theta}(b-x)\right]^k e^{-\sqrt{(2p+1)\theta}[(x-a)+(b-x)]}\,dx \\ +\int_{x=b}^{1}\left[\sqrt{(2p+1)\theta}(x-a)\right]^j\left[\sqrt{(2p+1)\theta}(x-b)\right]^k e^{-\sqrt{(2p+1)\theta}[(x-a)+(x-b)]}\,dx \end{matrix}\right\}.$$

As in Sec. 6, we proceed by changing variables.

For $x < a \le b$:

$$\tilde{x}_a \equiv \sqrt{(2p+1)\theta}(a-x),\ \ \tilde{x}_b \equiv \sqrt{(2p+1)\theta}(b-x),\ \ dx = \frac{-d\tilde{x}_a}{\sqrt{(2p+1)\theta}}, \tag{8.2}$$

$$\left[\sqrt{(2p+1)\theta}(a-x)\right]^j\left[\sqrt{(2p+1)\theta}(b-x)\right]^k = \tilde{x}_a^j\tilde{x}_b^k\,.$$

Solving the equations in Line 8.2 for $\tilde{x}_b$ in terms of $\tilde{x}_a$ gives

$$\tilde{x}_b = \sqrt{(2p+1)\theta}\left(b - a + \frac{\tilde{x}_a}{\sqrt{(2p+1)\theta}}\right) = \tilde{x}_a + \sqrt{(2p+1)\theta}(b-a) \text{ and}$$

$$\sqrt{(2p+1)\theta}[(a-x)+(b-x)] = \tilde{x}_a + \tilde{x}_b = 2\tilde{x}_a + \sqrt{(2p+1)\theta}(b-a). \tag{8.3}$$

For $a \le x \le b$, and in similar fashion:

$$\tilde{\tilde{x}}_a \equiv \sqrt{(2p+1)\theta}(x-a)\,,\ \tilde{\tilde{x}}_b = \sqrt{(2p+1)\theta}(b-x)\,,\ dx = \frac{d\tilde{\tilde{x}}_a}{\sqrt{(2p+1)\theta}}\,,$$

$$\left[\sqrt{(2p+1)\theta}(x-a)\right]^j\left[\sqrt{(2p+1)\theta}(b-x)\right]^k = \tilde{\tilde{x}}_a^j\tilde{\tilde{x}}_b^k\,,$$

$$\tilde{\tilde{x}}_b = \sqrt{(2p+1)\theta}\left(b - a - \frac{\tilde{\tilde{x}}_a}{\sqrt{(2p+1)\theta}}\right) = -\tilde{\tilde{x}}_a + \sqrt{(2p+1)\theta}(b-a) \text{ and}$$

$$\sqrt{(2p+1)\theta}[(x-a)+(b-x)] = \tilde{\tilde{x}}_a + \tilde{\tilde{x}}_b = \sqrt{(2p+1)\theta}(b-a). \tag{8.4}$$

For $a \le b < x$, and in similar fashion:

$$\tilde{\tilde{\tilde{x}}}_a \equiv \sqrt{(2p+1)\theta}(x-a)\,,\ \tilde{\tilde{\tilde{x}}}_b \equiv \sqrt{(2p+1)\theta}(x-b)\,,\ dx = \frac{d\tilde{\tilde{\tilde{x}}}_a}{\sqrt{(2p+1)\theta}}\,,$$

$$\left[\sqrt{(2p+1)\theta}(x-a)\right]^j\left[\sqrt{(2p+1)\theta}(x-b)\right]^k = \tilde{\tilde{\tilde{x}}}_a^j\tilde{\tilde{\tilde{x}}}_b^k\,,$$

$$\tilde{\tilde{\tilde{x}}}_b = \sqrt{(2p+1)\theta}\left(\frac{\tilde{\tilde{\tilde{x}}}_a}{\sqrt{(2p+1)\theta}} + a - b\right) = \tilde{\tilde{\tilde{x}}}_a - \sqrt{(2p+1)\theta}(b-a), \text{ and}$$

$$\sqrt{(2p+1)\theta}[(x-a)+(x-b)] = \tilde{\tilde{\tilde{x}}}_a + \tilde{\tilde{\tilde{x}}}_b = 2\tilde{\tilde{\tilde{x}}}_a - \sqrt{(2p+1)\theta}(b-a). \tag{8.5}$$

$$J_{p+1/2} = \frac{1}{\sqrt{(2p+1)\theta}}\left[\frac{p!}{(2p)!}\right]^2 \sum_{j,k=0}^{p} \frac{(2p-j)!(2p-k)!2^{j+k}}{2(p-j)!j!(p-k)!k!} \begin{pmatrix} -\int_{\tilde{x}_a=\sqrt{(2p+1)\theta}(1+a)}^{0} & \tilde{x}_a^j\tilde{x}_b^k e^{-(\tilde{x}_a+\tilde{x}_b)} d\tilde{x}_a \\ +\int_{\tilde{\tilde{x}}_a=0}^{\sqrt{(2p+1)\theta}(b-a)} & \tilde{\tilde{x}}_a^j\tilde{\tilde{x}}_b^k e^{-\left(\tilde{\tilde{x}}_a+\tilde{\tilde{x}}_b\right)}\, d\tilde{\tilde{x}}_a \\ +\int_{\tilde{\tilde{\tilde{x}}}_a=\sqrt{(2p+1)\theta}(b-a)}^{\sqrt{(2p+1)\theta}(1-a)} & \tilde{\tilde{\tilde{x}}}_a^j\tilde{\tilde{\tilde{x}}}_b^k e^{-\left(\tilde{\tilde{\tilde{x}}}_a+\tilde{\tilde{\tilde{x}}}_b\right)}\, d\tilde{\tilde{\tilde{x}}}_a \end{pmatrix}.$$

Then, using Eqs. 8.3, 8.4, and 8.5 to express $\tilde{x}_b$ (respectively, $\tilde{\tilde{x}}_b$ or $\tilde{\tilde{\tilde{x}}}_b$) in terms of $\tilde{x}_a$ (resp., $\tilde{\tilde{x}}_a$ or $\tilde{\tilde{\tilde{x}}}_a$), dropping all tildes and subscripts on the $x$'s, and making slight rearrangements, gives

$$J_{p+1/2} = \frac{1}{\sqrt{(2p+1)\theta}}\left[\frac{p!}{(2p)!}\right]^2 \sum_{j,k=0}^{p} \frac{(2p-j)!(2p-k)!2^{j+k}}{2(p-j)!j!(p-k)!k!}$$

$$\cdot\left\{\begin{array}{lll} -e^{-\sqrt{(2p+1)\theta}(b-a)}\int_{\sqrt{(2p+1)\theta}(1+a)}^{0} & x^j\left[x+\sqrt{(2p+1)\theta}(b-a)\right]^k e^{-2x} dx \\ + \ e^{-\sqrt{(2p+1)\theta}(b-a)}(-1)^k\int_{0}^{\sqrt{(2p+1)\theta}(b-a)} & x^j\left[x-\sqrt{(2p+1)\theta}(b-a)\right]^k \qquad dx \\ +e^{\sqrt{(2p+1)\theta}(b-a)}\int_{\sqrt{(2p+1)\theta}(b-a)}^{\sqrt{(2p+1)\theta}(1-a)} & x^j\left[x-\sqrt{(2p+1)\theta}(b-a)\right]^k e^{-2x}\, dx \end{array}\right\}.$$

Applying the binomial theorem, $(x \pm y)^n = \sum_{i=0}^{n}\frac{n!}{i!(n-i)!}(\pm 1)^i x^{n-i}y^i$, and factoring out $e^{-\sqrt{(2p+1)\theta}(b-a)}$ gives

$$J_{p+1/2} = \frac{1}{\sqrt{(2p+1)\theta}}\left[\frac{p!}{(2p)!}\right]^2 \sum_{j,k=0}^{p} \frac{(2p-j)!(2p-k)!2^{j+k}}{2(p-j)!j!(p-k)!k!} e^{-\sqrt{(2p+1)\theta}(b-a)}$$

$$\cdot\left\{\begin{array}{ll} -\sum_{l=0}^{k}\frac{k!}{l!(k-l)!}\int_{\sqrt{(2p+1)\theta}(1+a)}^{0} & x^{j+k-l}\left[\sqrt{(2p+1)\theta}(b-a)\right]^{l}e^{-2x}dx \\ +(-1)^{k+l}\sum_{l=0}^{k}\frac{k!}{l!(k-l)!}\int_{0}^{\sqrt{(2p+1)\theta}(b-a)} & x^{j+k-l}\left[\sqrt{(2p+1)\theta}(b-a)\right]^{l}\quad dx \\ +(-1)^{l}e^{2\sqrt{(2p+1)\theta}(b-a)}\sum_{l=0}^{k}\frac{k!}{l!(k-l)!}\int_{\sqrt{(2p+1)\theta}(b-a)}^{\sqrt{(2p+1)\theta}(1-a)} & x^{j+k-l}\left[\sqrt{(2p+1)\theta}(b-a)\right]^{l}e^{-2x}dx \end{array}\right\}.$$

Moving the summations and $\frac{k!}{l!(k-l)!}$ left, canceling the $k!$ in the numerator and denominator, and factoring $\left[\sqrt{(2p+1)\theta}(b-a)\right]^{l}$ gives

$$J_{p+1/2}=\frac{1}{\sqrt{(2p+1)\theta}}\left[\frac{p!}{(2p)!}\right]^{2}\sum_{j,k=0}^{p}\sum_{l=0}^{k}\frac{(2p-j)!(2p-k)!2^{j+k}}{2(p-j)!j!(p-k)!l!(k-l)!}e^{-\sqrt{(2p+1)\theta}(b-a)}\left[\sqrt{(2p+1)\theta}(b-a)\right]^{l}$$

$$\cdot\left[\begin{array}{r} -\int_{\sqrt{(2p+1)\theta}(1+a)}^{0}x^{j+k-l}e^{-2x}dx \\ +(-1)^{k+l}\int_{0}^{\sqrt{(2p+1)\theta}(b-a)}x^{j+k-l}\quad dx \\ +(-1)^{l}e^{2\sqrt{(2p+1)\theta}(b-a)}\int_{\sqrt{(2p+1)\theta}(b-a)}^{\sqrt{(2p+1)\theta}(1-a)}x^{j+k-l}e^{-2x}dx \end{array}\right]. \quad (8.6)$$

Substituting the integral of $x^{j+k-l}\,e^{-2x}$, Eq. 5.7, in the first and last rows of the large parentheses of Eq. 8.6, as well as the trivial integral $\int x^{j+k-l}dx=\frac{x^{j+k-l+1}}{j+k-l+1}$ in the second row of the large parentheses of Eq. 8.6, and some rearranging gives

$$J_{p+1/2}=\frac{1}{\sqrt{(2p+1)\theta}}\left[\frac{p!}{(2p)!}\right]^{2}\sum_{j,k=0}^{p}\sum_{l=0}^{k}\frac{(2p-j)!(2p-k)!2^{j+k}}{2(p-j)!j!(p-k)!l!(k-l)!}\left[\sqrt{(2p+1)\theta}(b-a)\right]^{l}e^{-\sqrt{(2p+1)\theta}(b-a)}$$

$$\cdot\left[\begin{array}{r} \frac{(j+k-l)!}{2^{j+k-l+1}}e^{-2x}\sum_{m=0}^{j+k-l}\frac{(2x)^{m}}{m!}\Big|_{\sqrt{(2p+1)\theta}(1+a)}^{0} \\ +(-1)^{k+l}\frac{x^{j+k-l+1}}{(j+k-l+1)}\Big|_{x=0}^{\sqrt{(2p+1)\theta}(b-a)} \\ -(-1)^{l}\frac{(j+k-l)!}{2^{j+k-l+1}}e^{2\sqrt{(2p+1)\theta}(b-a)}e^{-2x}\sum_{m=0}^{j+k-l}\frac{(2x)^{m}}{m!}\Big|_{\sqrt{(2p+1)\theta}(b-a)}^{\sqrt{(2p+1)\theta}(1-a)} \end{array}\right]. \quad (8.7)$$

Evaluation of the integrals gives the following, where the identity $0^{0}=1$ has been used in the evaluation of the upper limit of the first integral in Eq. 8.7, the double-factorial identity of Eq. 5.1 has been used, and $\frac{(j+k-l)!}{2^{j+k-l+1}}$ has been factored out. Each row is given a different color, as this will prove useful in what follows.

$$J_{p+1/2}=\frac{1}{\sqrt{(2p+1)\theta}4^{p}[(2p-1)!!]^{2}}\sum_{j,k=0}^{p}\sum_{l=0}^{k}\frac{(2p-j)!(2p-k)!2^{j+k}}{2(p-j)!j!(p-k)!l!(k-l)!}\left[\sqrt{(2p+1)\theta}(b-a)\right]^{l}e^{-\sqrt{(2p+1)\theta}(b-a)}$$

$$\cdot\frac{(j+k-l)!}{2^{j+k-l+1}}\left\{\begin{array}{c} 1 \\ -\sum_{m=0}^{j+k-l}\frac{1}{m!}\left[2\sqrt{(2p+1)\theta}(1+a)\right]^{m} e^{-2\sqrt{(2p+1)\theta}(1+a)} \\ +(-1)^{k+l}\frac{1}{(j+k-l+1)!}\quad\left[2\sqrt{(2p+1)\theta}(b-a)\right]^{j+k-l+1} \\ -(-1)^{l}\sum_{m=0}^{j+k-l}\frac{1}{m!}\left[2\sqrt{(2p+1)\theta}(1-a)\right]^{m} e^{-2\sqrt{(2p+1)\theta}(1-a)}e^{2\sqrt{(2p+1)\theta}(b-a)} \\ +(-1)^{l}\sum_{m=0}^{j+k-l}\frac{1}{m!}\left[2\sqrt{(2p+1)\theta}(b-a)\right]^{m} e^{-2\sqrt{(2p+1)\theta}(b-a)}e^{2\sqrt{(2p+1)\theta}(b-a)} \end{array}\right\}. \tag{8.8}$$

Using the identities

$$e^{-\sqrt{(2p+1)\theta}(b-a)}e^{-2\sqrt{(2p+1)\theta}(1+a)} = e^{-2\sqrt{(2p+1)\theta}\left(1+\frac{a+b}{2}\right)} \text{ and}$$

$$e^{\sqrt{(2p+1)\theta}(b-a)}e^{-2\sqrt{(2p+1)\theta}(1-a)} = e^{-2\sqrt{(2p+1)\theta}\left(1-\frac{a+b}{2}\right)},$$

collecting terms with common exponents, moving the factor $\frac{(j+k-l)!}{2^{j+k-l+1}}$ left, distributing $\left[\sqrt{(2p+1)\theta}(b-a)\right]^{l}$, using the elementary identity $\frac{2^{j+k}}{2^{j+k-l+1}} = \frac{2^{l}}{2}$, and maintaining the colors of Eq. 8.8 to readily demonstrate the rearrangement gives

$$J_{p+1/2} = \frac{1}{\sqrt{(2p+1)\theta}}\left[\frac{p!}{(2p)!}\right]^{2}\sum_{j,k=0}^{p}\sum_{l=0}^{k}\frac{(2p-j)!(2k-k)!2^{l}(j+k-l)!}{4(p-j)!j!(p-k)!l!(k-l)!}\left[\sqrt{(2p+1)\theta}(b-a)\right]^{l}$$

$$\cdot\left(\begin{array}{c} \left\{\begin{array}{c} 1 \\ +(-1)^{k+l}\frac{1}{(j+k-l+1)!}\left[2\sqrt{(2p+1)\theta}(b-a)\right]^{j+k-l+1} \\ +(-1)^{l}\sum_{m=0}^{j+k-l}\frac{1}{m!}\left[2\sqrt{(2p+1)\theta}(b-a)\right]^{m} \end{array}\right\} e^{-2\sqrt{(2p+1)\theta}\left(\frac{b-a}{2}\right)} \\ -\left\{\sum_{\{=0}^{j+k-l}\frac{1}{m!}\left[2\sqrt{(2p+1)\theta}(1+a)\right]^{m}\right\} e^{-2\sqrt{(2p+1)\theta}\left(1+\frac{a+b}{2}\right)} \\ +\left\{(-1)^{l}\sum_{m=0}^{j+k-l}\frac{1}{\{!}\left[2\sqrt{(2p+1)\theta}(1-a)\right]^{\{}\right\} e^{-2\sqrt{(2p+1)\theta}\left(1-\frac{a+b}{2}\right)} \end{array}\right). \tag{8.9}$$

Consolidating the $2^{l}$ and the $\left[\sqrt{(2p+1)\theta}(b-a)\right]^{l}$, and using the assumption $b \geq a$, gives

$$J_{p+1/2} = \frac{1}{\sqrt{(2p+1)\theta}}\left[\frac{p!}{(2p)!}\right]^{2}\sum_{j,k=0}^{p}\sum_{l=0}^{k}\frac{(2p-j)!(2p-k)!(j+k-l)!}{4(p-j)!j!(p-k)!l!(k-l)!}\left[2\sqrt{(2p+1)\theta}\,|b-a|\right]^{l}$$

$$\cdot\left(\begin{array}{c} \left\{\begin{array}{c} 1 \\ +(-1)^{k+l}\frac{1}{(j+k-l+1)!}\left[2\sqrt{(2p+1)\theta}|b-a|\right]^{j+k-l+1} \\ +(-1)^{l}\sum_{m=0}^{j+k-l}\frac{1}{m!}\left[2\sqrt{(2p+1)\theta}|b-a|\right]^{m} \end{array}\right\} e^{-2\sqrt{(2p+1)\theta}\left|\frac{b-a}{2}\right|} \\ -\left\{\sum_{m=0}^{j+k-l}\frac{1}{m!}\left[2\sqrt{(2p+1)\theta}(1+a)\right]^{m}\right\} e^{-2\sqrt{(2p+1)\theta}\left(1+\frac{a+b}{2}\right)} \\ +\left\{(-1)^{l}\sum_{m=0}^{j+k-l}\frac{1}{m!}\left[2\sqrt{(2p+1)\theta}(1-a)\right]^{m}\right\} e^{-2\sqrt{(2p+1)\theta}\left(1-\frac{a+b}{2}\right)} \end{array}\right),\quad b \geq a. \tag{8.10}$$

Introducing the definitions

$$A_{\pm} \equiv 2\sqrt{(2p+1)\theta}\,(1 \pm a) \text{ and} \tag{8.11}$$

$$B \equiv 2\sqrt{(2p+1)\theta}\,|b-a| \tag{8.12}$$

gives the following alternative equation that carries through the sums over $m$:

$$J_{p+1/2} = \frac{1}{\sqrt{(2p+1)\theta}}\left[\frac{p!}{(2p)!}\right]^2 \sum_{j,k=0}^{p}\sum_{l=0}^{k}\frac{(2p-j)!(2p-k)!(j+k-l)!}{4(p-j)!j!(p-k)!l!(k-l)!}$$

$$\cdot\left(\begin{array}{r}\left\{B^l + (-1)^l B^l\left[1 + B + \frac{B^2}{2} + \cdots + \frac{B^{j+k-l}}{(j+k-l)!}\right] + \frac{(-1)^{k+l}B^{j+k+1}}{(j+k-l+1)!}\right\} e^{-2\sqrt{(2p+1)\theta}\left|\frac{b-a}{2}\right|} \\ -B^l\left\{1 + A_+ + \frac{A_+^2}{2} + \cdots + \frac{A_+^{j+k-l}}{(j+k-l)!}\right\} e^{-2\sqrt{(2p+1)\theta}\left(1+\frac{a+b}{2}\right)} \\ +(-1)^l B^l\left\{1 + A_- + \frac{A_-^2}{2} + \cdots + \frac{A_-^{j+k-l}}{(j+k-l)!}\right\} e^{-2\sqrt{(2p+1)\theta}\left(1-\frac{a+b}{2}\right)}\end{array}\right), \quad b \geq a. \tag{8.13}$$

Finally, identifying $J_{p+1/2}$, $a$, and $b$ of Eq. 8.13 with $R_{i,j}^{(p)}$, $x_{i,k}$, and $x_{j,k}$, respectively, provides the following two alternative expressions that can replace the “machine-readable symbolic expressions,” i.e., the “MRSE’s,” in Part I, Table 4.3 [1]:

$$R_{i,j}^{(p)} = \frac{1}{\sqrt{(2p+1)\theta}}\left[\frac{p!}{(2p)!}\right]^2 \sum_{j,k=0}^{p}\sum_{l=0}^{k}\frac{(2p-j)!(2p-k)!(j+k-l)!}{4(p-j)!j!(p-k)!l!(k-l)!}\left[2\sqrt{(2p+1)\theta}\,|b-a|\right]^l$$

$$\cdot\left(\begin{array}{l}\left\{\begin{array}{c}1 \\ +(-1)^{k+l}\frac{1}{(j+k-l+1)!}\left[2\sqrt{(2p+1)\theta}|b-a|\right]^{j+k-l+1} \\ +(-1)^l\sum_{m=0}^{j+k-l}\frac{1}{m!}\left[2\sqrt{(2p+1)\theta}|b-a|\right]^m\end{array}\right\} e^{-2\sqrt{(2p+1)\theta}\left|\frac{x_{j,k}-x_{i,k}}{2}\right|} \\ -\left\{\sum_{m=0}^{j+k-l}\frac{1}{m!}\left[2\sqrt{(2p+1)\theta}(1+a)\right]^m\right\} e^{-2\sqrt{(2p+1)\theta}\left(1+\frac{x_{j,k}+x_{i,k}}{2}\right)} \\ +\left\{(-1)^l\sum_{m=0}^{j+k-l}\frac{1}{m!}\left[2\sqrt{(2p+1)\theta}(1-a)\right]^m\right\} e^{-2\sqrt{(2p+1)\theta}\left(1-\frac{x_{j,k}+x_{i,k}}{2}\right)}\end{array}\right), \quad x_{j,k} \geq x_{i,k}, \tag{8.14}$$

or

$$R_{i,j}^{(p)} = \frac{1}{\sqrt{(2p+1)\theta}}\left[\frac{p!}{(2p)!}\right]^2 \sum_{j,k=0}^{p}\sum_{l=0}^{k}\frac{(2p-j)!(2p-k)!(j+k-l)!}{4(p-j)!j!(p-k)!l!(k-l)!}$$

$$\cdot\left(\begin{array}{r}\left\{B^l + (-1)^l B^l\left[1 + B + \frac{B^2}{2} + \cdots + \frac{B^{j+k-l}}{(j+k-l)!}\right] + \frac{(-1)^{k+l}B^{j+k+1}}{(j+k-l+1)!}\right\} e^{-2\sqrt{(2p+1)\theta}\left|\frac{x_{j,k}-x_{i,k}}{2}\right|} \\ -B^l\left\{1 + A_+ + \frac{A_+^2}{2} + \cdots + \frac{A_+^{j+k-l}}{(j+k-l)!}\right\} e^{-2\sqrt{(2p+1)\theta}\left(1+\frac{x_{j,k}+x_{i,k}}{2}\right)} \\ +(-1)^l B^l\left\{1 + A_- + \frac{A_-^2}{2} + \cdots + \frac{A_-^{j+k-l}}{(j+k-l)!}\right\} e^{-2\sqrt{(2p+1)\theta}\left(1-\frac{x_{j,k}+x_{i,k}}{2}\right)}\end{array}\right), \quad x_{j,k} \geq x_{i,k}. \tag{8.15}$$

## 9. Two example integrals of products of two Matérn-correlation functions: $J_{3/2}$ and $J_{5/2}$

Each example of this section uses Eq. 8.15.

**Example:** $\boldsymbol{p=1}$**:** $\boldsymbol{J_{\nu=p+1/2} \equiv J_{3/2} \equiv R_{i,j}^{(p=1)}},\ \ \boldsymbol{1 \le i, j \le}$ **n**.

$$J_{3/2} = \frac{1}{16\sqrt{3\theta}} \sum_{j,k=0}^{1} \sum_{l=0}^{k} (2-j)!\,(2-k)!\,(j+k-l)!$$

$$\cdot \left( \begin{array}{r} \left\{ B^l + (-1)^l B^l \left[ 1 + B + \frac{B^2}{2} + \cdots + \frac{B^{j+k-l}}{(j+k-l)!} \right] + \frac{(-1)^{k+l} B^{j+k+1}}{(j+k-l+1)!} \right\} e^{-2\sqrt{3\theta}\left(\frac{b-a}{2}\right)} \\ -B^l \left\{ 1 + A_+ + \frac{A_+^2}{2} + \cdots + \frac{A_+^{j+k-l}}{(j+k-l)!} \right\} e^{-2\sqrt{3\theta}\left(1+\frac{a+b}{2}\right)} \\ +(-1)^l B^l \left\{ 1 + A_- + \frac{A_-^2}{2} + \cdots + \frac{A_-^{j+k-l}}{(j+k-l)!} \right\} e^{-2\sqrt{3\theta}\left(1-\frac{a+b}{2}\right)} \end{array} \right).$$

Terms in the last equation's large parentheses are evaluated individually for each $[j, k, l]$ set, to give the sub-terms of $J_{3/2}$, each with an integer 16 in its denominator, as follows:

$$[0,0,0]: \frac{4}{16\sqrt{3\theta}} \begin{bmatrix} (2+B) e^{-2\sqrt{3\theta}\left(\frac{b-a}{2}\right)} \\ -(1) e^{-2\sqrt{3\theta}\left(1+\frac{a+b}{2}\right)} \\ +(1) e^{-2\sqrt{3\theta}\left(1-\frac{a+b}{2}\right)} \end{bmatrix}. \qquad [0,1,0]: \frac{2}{16\sqrt{3\theta}} \begin{bmatrix} \left(2+B-\frac{1}{2}B^2\right) e^{-2\sqrt{3\theta}\left(\frac{b-a}{2}\right)} \\ -(1+A_+) e^{-2\sqrt{3\theta}\left(1+\frac{a+b}{2}\right)} \\ +(1+A_-) e^{-2\sqrt{3\theta}\left(1-\frac{a+b}{2}\right)} \end{bmatrix}.$$

$$[0,1,1]: \frac{2}{16\sqrt{3\theta}} \begin{bmatrix} (B^2) e^{-2\sqrt{3\theta}\left(\frac{b-a}{2}\right)} \\ -B(1) e^{-2\sqrt{3\theta}\left(1+\frac{a+b}{2}\right)} \\ +B(-1) e^{-2\sqrt{3\theta}\left(1-\frac{a+b}{2}\right)} \end{bmatrix}. \qquad [1,0,0]: \frac{2}{16\sqrt{3\theta}} \begin{bmatrix} \left(2+B+\frac{1}{2}B^2\right) e^{-2\sqrt{3\theta}\left(\frac{b-a}{2}\right)} \\ -(1+A_+) e^{-2\sqrt{3\theta}\left(1+\frac{a+b}{2}\right)} \\ +(1+A_-) e^{-2\sqrt{3\theta}\left(1-\frac{a+b}{2}\right)} \end{bmatrix}.$$

$$[1,1,0]: \frac{2}{16\sqrt{3\theta}} \begin{bmatrix} \left(2+B+\frac{1}{2}B^2-\frac{1}{6}B^3\right) e^{-2\sqrt{3\theta}\left(\frac{b-a}{2}\right)} \\ -\left(1+A_++\frac{1}{2}A_+^2\right) e^{-2\sqrt{3\theta}\left(1+\frac{a+b}{2}\right)} \\ +\left(1+A_-+\frac{1}{2}A_-^2\right) e^{-2\sqrt{3\theta}\left(1-\frac{a+b}{2}\right)} \end{bmatrix}. \quad [1,1,1]: \frac{1}{16\sqrt{3\theta}} \begin{bmatrix} \left(-B^2+\frac{1}{2}B^3\right) e^{-2\sqrt{3\theta}\left(\frac{b-a}{2}\right)} \\ -B(1+A_+) e^{-2\sqrt{3\theta}\left(1+\frac{a+b}{2}\right)} \\ +B(-1-A_-) e^{-2\sqrt{3\theta}\left(1-\frac{a+b}{2}\right)} \end{bmatrix}.$$

Summing these six terms gives

$$J_{3/2} = \frac{1}{16\sqrt{3\theta}} \left\{ \begin{array}{c} \left[20 + 10B + 2B^2 + \frac{1}{6}B^3\right] e^{-2\sqrt{3\theta}\left(\frac{b-a}{2}\right)} \\ -[(10 + 6A_+ + A_+^2) + B(3 + A_+)] e^{-2\sqrt{3\theta}\left(1+\frac{a+b}{2}\right)} \\ +[(10 + 6A_- + A_-^2) - B(3 + A_-)] e^{-2\sqrt{3\theta}\left(1-\frac{a+b}{2}\right)} \end{array} \right\}.$$

Then, changing the denominator factor to 24 for comparison with previous results in Parts I and Ia of this series of papers, gives

$$J_{3/2} = \frac{1}{24\sqrt{3\theta}} \left\{ \begin{array}{c} \left[30 + 15B + 3B^2 + \frac{1}{4}B^3\right] e^{-2\sqrt{3\theta}\left(\frac{b-a}{2}\right)} \\ -\left[\left(15 + 9A_+ + \frac{3}{2}A_+^2\right) + B\left(\frac{9}{2} + \frac{3}{2}A_+\right)\right] e^{-2\sqrt{3\theta}\left(1+\frac{a+b}{2}\right)} \\ +\left[\left(15 + 9A_- + \frac{3}{2}A_-^2\right) - B\left(\frac{9}{2} + \frac{3}{2}A_-\right)\right] e^{-2\sqrt{3\theta}\left(1-\frac{a+b}{2}\right)} \end{array} \right\}.$$

Substituting back in the definitions of $A_\pm$ and $B$ in Eqs. 8.11 and 8.12 gives

$$J_{3/2} = \frac{1}{24\sqrt{3\theta}} \left( \begin{array}{c} \left\{ \begin{array}{l} 30 \\ +30\sqrt{3\theta}\ (b-a) \\ +12 \cdot 3\theta\ (b-a)^2 \\ +\ 2\sqrt{3\theta}^3 (b-a)^3 \end{array} \right\} e^{-2\sqrt{3\theta}\left(\frac{b-a}{2}\right)} \\ -\left\{ \left[ \begin{array}{l} 15 \\ +18\sqrt{3\theta}(1+a) \\ +\ \ 6 \cdot 3\theta(1+a)^2 \end{array} \right] + \sqrt{3\theta}(b-a) \left[ \begin{array}{l} 9 \\ +6\sqrt{3\theta}(1+a) \end{array} \right] \right\} e^{-2\sqrt{3\theta}\left(1+\frac{a+b}{2}\right)} \\ +\left\{ \left[ \begin{array}{l} 15 \\ +18\sqrt{3\theta}(1-a) \\ +\ \ 6 \cdot 3\theta(1-a)^2 \end{array} \right] - \sqrt{3\theta}(b-a) \left( \begin{array}{l} 9 \\ +6\sqrt{3\theta}(1-a) \end{array} \right) \right\} e^{-2\sqrt{3\theta}\left(1-\frac{a+b}{2}\right)} \end{array} \right)$$

$$= \frac{1}{24\sqrt{3\theta}} \left\{ \begin{array}{c} \left[ \begin{array}{l} 30 \\ +30\sqrt{3\theta}(b-a) \\ +12 \cdot 3\theta(b-a)^2 \\ +2\sqrt{3\theta}^3 (b-a)^3 \end{array} \right] e^{-2\sqrt{3\theta}\left(\frac{b-a}{2}\right)} \\ -\left[ \begin{array}{l} 15 \\ +\ \ 9\sqrt{3\theta}(2+a+b) \\ +\ \ 6 \cdot 3\theta(1+a+b+ab) \end{array} \right] e^{-2\sqrt{3\theta}\left(1+\frac{a+b}{2}\right)} \\ +\left[ \begin{array}{l} 15 \\ +\ \ 9\sqrt{3\theta}(2-a-b) \\ +\ \ 6 \cdot 3\theta(1-a-b+ab) \end{array} \right] e^{-2\sqrt{3\theta}\left(1-\frac{a+b}{2}\right)} \end{array} \right\}.$$

Factoring some more gives

$$J_{3/2} = \frac{1}{24\sqrt{3\theta_k}} \left\{ \begin{array}{c} 2\left[ \begin{array}{l} 15 \\ +15\left|x_{i,k} - x_{j,k}\right|\ \sqrt{3\theta_k} \\ +\ \ 6\left|x_{i,k} - x_{j,k}\right|^2 \sqrt{3\theta_k}^2 \\ +\ \ \ \left|x_{i,k} - x_{j,k}\right|^3 \sqrt{3\theta_k}^3 \end{array} \right] e^{-2\sqrt{3\theta_k}\left|\frac{x_{i,k}-x_{j,k}}{2}\right|} \\ -3\mathcal{T}^{(-)}_{x_{i,k};x_{j,k}} \left[ \begin{array}{l} 5 \\ +3\left( \begin{array}{l} 2 \\ +x_{i,k} + x_{j,k} \end{array} \right) \sqrt{3\theta_k} \\ +2\left( \begin{array}{l} 1 \\ +x_{i,k} + x_{j,k} \\ +x_{i,k}x_j, k \end{array} \right) \sqrt{3\theta_k}^2 \end{array} \right] e^{-2\sqrt{3\theta_k}\left(1+\frac{x_{i,k}+x_{j,k}}{2}\right)} \end{array} \right\}. \quad (9.1)$$

This agrees with the result for this integral in Paper Ia, Sec. 4 of this series of papers [2], viz.,

$$R_{i,j}^{(1)} = \frac{1}{24\sqrt{3\theta_k}} \left\{ \begin{array}{r} 2\begin{bmatrix} 15 \\ +15|x_{i,k} - x_{j,k}| \sqrt{3\theta_k} \\ + \quad 6|x_{i,k} - x_{j,k}|^2 \sqrt{3\theta_k}^2 \\ + \quad |x_{i,k} - x_{j,k}|^3 \sqrt{3\theta_k}^3 \end{bmatrix} e^{-2\sqrt{3\theta_k}\left|\frac{x_{i,k}-x_{j,k}}{2}\right|} \\ -3\mathcal{T}_{x_{i,k};x_{j,k}}^{(-)} \begin{bmatrix} 5 \\ +3\begin{pmatrix} 2 \\ +x_{i,k} + x_{j,k} \end{pmatrix} \sqrt{3\theta_k}^1 \\ +2\begin{pmatrix} 1 \\ +x_{i,k} + x_{j,k} \\ +x_{i,k}x_{j,k} \end{pmatrix} \sqrt{3\theta_k}^2 \end{bmatrix} e^{-2\sqrt{3\theta_k}\left(1+\frac{x_{i,k}+x_{j,k}}{2}\right)} \end{array} \right\}.$$

**Example:** ***p*=2:** $\boldsymbol{J_{\nu=p+1/2} \equiv J_{5/2} \equiv R_{i,j}^{(p=2)}}$, $\boldsymbol{1 \le i, j \le n}$.

$$J_{5/2} = \frac{1}{576\sqrt{5\theta}} \sum_{j,k=0}^{2} \sum_{l=0}^{k} \frac{(4-j)!(4-k)!(j+k-l)!}{(2-j)!j!(2-k)!l!(k-l)!}$$

$$\cdot \left( \begin{array}{r} \left\{B^l + (-1)^l B^l \left[1 + B + \frac{B^2}{2} + \cdots + \frac{B^{j+k-l}}{(j+k-l)!}\right] + \frac{(-1)^{k+l} B^{j+k+1}}{(j+k-l+1)!}\right\} e^{-2\sqrt{5\theta}\left(\frac{b-a}{2}\right)} \\ -B^l \left\{1 + A_+ + \frac{A_+^2}{2} + \cdots + \frac{A_+^{j+k-l}}{(j+k-l)!}\right\} e^{-2\sqrt{5\theta}\left(1+\frac{a+b}{2}\right)} \\ +(-1)^l B^l \left\{1 + A_- + \frac{A_-^2}{2} + \cdots + \frac{A_-^{j+k-l}}{(j+k-l)!}\right\} e^{-2\sqrt{5\theta}\left(1-\frac{a+b}{2}\right)} \end{array} \right).$$

Terms in the last equation's large parentheses are evaluated individually for each $[j, k, l]$ set, to give the following sub-terms of $J_{5/2}$, each with an integer 576 in its denominator, with the help of the definitions

$$E_\delta \equiv e^{-2\sqrt{5\theta}\left(\frac{b-a}{2}\right)} \text{ and} \tag{9.2}$$

$$E_\pm \equiv e^{-2\sqrt{5\theta}\left(1\pm\frac{a+b}{2}\right)}: \tag{9.3}$$

$$[0,0,0]: \frac{144}{576\sqrt{5\theta}}\begin{bmatrix}(2+B)E_\delta\\ -(1)E_+\\ +(1)E_-\end{bmatrix}$$

$$[0,1,0]: \frac{72}{576\sqrt{5\theta}}\begin{bmatrix}\left(2+B-\frac{1}{2}B^2\right)E_\delta\\ -(1+A_+)E_+\\ +(1+A_-)E_-\end{bmatrix}$$

$$[0,1,1]: \frac{72}{576\sqrt{5\theta}}\begin{bmatrix}(B^2)E_\delta\\ -B(1)E_+\\ +B(-1)E_-\end{bmatrix}$$

$$[1,0,0]: \frac{72}{576\sqrt{5\theta}}\begin{bmatrix}\left(2+B+\frac{1}{2}B^2\right)E_\delta\\ -(1+A_+)E_+\\ +(1+A_-)E_-\end{bmatrix}$$

$$[1,1,0]: \frac{72}{576\sqrt{5\theta}}\begin{bmatrix}\left(2+B+\frac{1}{2}B^2-\frac{1}{6}B^3\right)E_\delta\\ -\left(1+A_++\frac{1}{2}A_+^2\right)E_+\\ +\left(1+A_-+\frac{1}{2}A_-^2\right)E_-\end{bmatrix}$$

$$[1,1,1]: \frac{36}{576\sqrt{5\theta}}\begin{bmatrix}\left(-B^2+\frac{1}{2}B^3\right)E_\delta\\ -B(1+A_+)E_+\\ +B(-1-A_-)E_-\end{bmatrix}$$

$$[0,2,0]: \frac{24}{576\sqrt{5\theta}}\begin{bmatrix}\left(2+B+\frac{1}{2}B^2+\frac{1}{6}B^3\right)E_\delta\\ -\left(1+A_++\frac{1}{2}A_+^2\right)E_+\\ +\left(1+A_-+\frac{1}{2}A_-^2\right)E_-\end{bmatrix}$$

$$[0,2,1]: \frac{24}{576\sqrt{5\theta}}\begin{bmatrix}\left(-B^2-\frac{1}{2}B^3\right)E_\delta\\ -B(1+A_+)E_+\\ +B(-1-A_-)E_-\end{bmatrix}$$

$$[0,2,2]: \frac{12}{576\sqrt{5\theta}}\begin{bmatrix}(2B^2+B^3)E_\delta\\ -B^2(1)E_+\\ +B^2(1)E_-\end{bmatrix}$$

$$[1,2,0]: \frac{36}{576\sqrt{5\theta}}\begin{bmatrix}\left(2+B+\frac{1}{2}B^2+\frac{1}{6}B^3+\frac{1}{24}B^4\right)E_\delta\\ -\left(1+A_++\frac{1}{2}A_+^2+\frac{1}{6}A_+^3\right)E_+\\ +\left(1+A_-+\frac{1}{2}A_-^2+\frac{1}{6}A_-^3\right)E_-\end{bmatrix}$$

$$[1,2,1]: \frac{24}{576\sqrt{5\theta}}\begin{bmatrix}\left(-B^2-\frac{1}{2}B^3-\frac{1}{6}B^4\right)E_\delta\\ -B\left(1+A_++\frac{1}{2}A_+^2\right)E_+\\ +B\left(-1-A_--\frac{1}{2}A_-^2\right)E_-\end{bmatrix}$$

$$[1,2,2]: \frac{6}{576\sqrt{5\theta}}\begin{bmatrix}\left(2B^2+B^3+\frac{1}{2}B^4\right)E_\delta\\ -B^2(1+A_+)E_+\\ +B^2(1+A_-)E_-\end{bmatrix}$$

$$[2,0,0]: \frac{24}{576\sqrt{5\theta}}\begin{bmatrix}\left(2+B+\frac{1}{2}B^2+\frac{1}{6}B^3\right)E_\delta\\ -\left(1+A_++\frac{1}{2}A_+^2\right)E_+\\ +\left(1+A_-+\frac{1}{2}A_-^2\right)E_-\end{bmatrix}$$

$$[2,1,0]: \frac{36}{576\sqrt{5\theta}}\begin{bmatrix}\left(2+B+\frac{1}{2}B^2+\frac{1}{6}B^3-\frac{1}{24}B^4\right)E_\delta\\ -\left(1+A_++\frac{1}{2}A_+^2+\frac{1}{6}A_+^3\right)E_+\\ +\left(1+A_-+\frac{1}{2}A_-^2+\frac{1}{6}A_-^3\right)E_-\end{bmatrix}$$

$$[2,1,1]: \frac{12}{576\sqrt{5\theta}}\begin{bmatrix}\left(-B^2-\frac{1}{2}B^3+\frac{1}{6}B^4\right)E_\delta\\ -B\left(1+A_++\frac{1}{2}A_+^2\right)E_+\\ +B\left(-1-A_--\frac{1}{2}A_-^2\right)E_-\end{bmatrix}$$

$$[2,2,0]: \frac{24}{576\sqrt{5\theta}}\begin{bmatrix}\left(2+B+\frac{B^2}{2}+\frac{B^3}{6}+\frac{B^4}{24}+\frac{B^5}{120}\right)E_\delta\\ -\left(1+A_++\frac{A_+^2}{2}+\frac{A_+^3}{6}+\frac{A_+^4}{24}\right)E_+\\ +\left(1+A_-+\frac{A_-^2}{2}+\frac{A_-^3}{6}+\frac{A_-^4}{24}\right)E_-\end{bmatrix}$$

$$[2,2,1]: \frac{12}{576\sqrt{5\theta}}\begin{bmatrix}\left(-B^2-\frac{1}{2}B^3-\frac{1}{6}B^4-\frac{1}{24}B^5\right)E_\delta\\ -B\left(1+A_++\frac{1}{2}A_+^2+\frac{1}{6}A_+^3\right)E_+\\ +B\left(-1-A_--\frac{1}{2}A_-^2-\frac{1}{6}A_-^3\right)E_-\end{bmatrix}$$

$$[2,2,2]: \frac{2}{576\sqrt{5\theta}}\begin{bmatrix}\left(2B^2+B^3+\frac{1}{2}B^4+\frac{1}{6}B^5\right)E_\delta\\ -B^2\left(1+A_++\frac{1}{2}A_+^2\right)E_+\\ +B^2\left(1+A_-+\frac{1}{2}A_-^2\right)E_-\end{bmatrix}.$$

Summing these eighteen terms gives

$$J_{5/2}=\frac{1}{576\sqrt{5\theta}}\left(\begin{array}{l}\left\{\begin{array}{l}\quad(288+144+144+144+48+72+48+72+48)\\+B\;\;(144+72+72+72+24+36+24+36+24)\\+B^2\begin{pmatrix}-36+72+36+36-36+12-24+24\\+18-24+12+12+18-12+12-12+4\end{pmatrix}\\+B^3(-12+18+4-12+12+6-12+6+4+6-6+4-6+2)\\+B^4\left(\frac{3}{2}-4+3-\frac{3}{2}+2+1-2+1\right)\\+B^5\left(\frac{1}{5}-\frac{1}{2}+\frac{1}{3}\right)\end{array}\right\}E_\delta\\-\left\{\begin{array}{l}\left[\begin{array}{l}(144+72+72+72+24+36+24+36+24)\\+(72+72+72+24+36+24+36+24)A_+\\+(36+12+18+12+18+12)A_+^2\\+(6+6+4)A_+^3+A_+^4\end{array}\right]\\+B\left[\begin{array}{l}(72+36+24+24+12+12)\\+(36+24+24+12+12)A_+\\+(12+6+6)A_+^2+2A_+^3\end{array}\right]\\+B^2[(12+6+2)+(6+2)A_++A_+^2]\end{array}\right\}E_+\\+\left\{\begin{array}{l}\left[\begin{array}{l}(144+72+72+72+24+36+24+36+24)\\+(72+72+72+24+36+24+36+24)A_-\\+(36+12+18+12+18+12)A_-^2\\+(6+6+4)A_-^3+A_-^4\end{array}\right]\\-B\left[\begin{array}{l}(72+36+24+24+12+12)\\+(36+24+24+24+12+12)A_-\\+(12+6+6)A_-^2+2A_-^3\end{array}\right]\\+B^2[(12+6+2)+(6+2)A_-+A_-^2]\end{array}\right\}E_-\end{array}\right)$$

$$=\frac{1}{576\sqrt{5\theta}}\left\{\begin{array}{l}\left[1008+504B+112B^2+14B^3+B^4+\frac{1}{30}B^5\right]E_\delta\\-\left[\begin{array}{l}(504+360A_++108A_+^2+16A_+^3+A_+^4)\\+B(180+108A_++24A_+^2+2A_+^3)+B^2(20+8A_++A_+^2)\end{array}\right]E_+\\+\left[\begin{array}{l}(504+360A_++108A_+^2+16A_+^3+A_+^4)\\-B(180+132A_++24A_+^2+2A_+^3)+B^2(24+8A_++A_+^2)\end{array}\right]E_-\end{array}\right\}.$$

Multiplying both numerator and denominator by $\frac{15}{8}$ gives a denominator of 1080 for comparison with previous results in Parts I and Ia of this series of papers:

$$J_{5/2}=\frac{1}{1080\sqrt{5\theta}}\left\{\begin{array}{l}\left[1890+945B+210B^2+\frac{105}{4}B^3+\frac{15}{8}B^4+\frac{1}{16}B^5\right]E_\delta\\-\left[\begin{array}{l}\left(945+675A_++\frac{405}{2}A_+^2+30A_+^3+\frac{15}{8}A_+^4\right)\\+B\left(\frac{675}{2}+\frac{405}{2}A_++45A_+^2+\frac{15}{4}A_+^3\right)+B^2\left(\frac{75}{2}+15A_++\frac{15}{8}A_+^2\right)\end{array}\right]E_+\\+\left[\begin{array}{l}\left(945+675A_-+\frac{405}{2}A_-^2+30A_-^3+\frac{15}{8}A_-^4\right)\\-B\left(\frac{675}{2}+\frac{495}{2}A_-+45A_-^2+\frac{15}{4}A_-^3\right)+B^2\left(45+15A_-+\frac{15}{8}A_-^2\right)\end{array}\right]E_-\end{array}\right\}.$$

Substituting back in the definitions of $A_{\pm}$ and $B$ of Eqs. 8.11 and 8.12, gives

$$J_{5/2}=\frac{1}{1080\sqrt{5\theta}}\left(\begin{array}{c}
2\left\{\begin{array}{c}945+945(b-a)\sqrt{5\theta}+420(b-a)^2\sqrt{5\theta}^2\\ +105(b-a)^3\sqrt{5\theta}^3+15(b-a)^4\sqrt{5\theta}^4+(b-a)^5\sqrt{5\theta}^5\end{array}\right\}E_\delta\\
-\left\{\begin{array}{l}\left[\begin{array}{c}945+1350(1+a)\sqrt{5\theta}+810(1+a)^2\sqrt{5\theta}^2\\ +240(1+a)^3\sqrt{5\theta}^3+30(1+a)^4\sqrt{5\theta}^4\end{array}\right]\\ +(b-a)\sqrt{5\theta}\left[\begin{array}{c}675+810(1+a)\sqrt{5\theta}\\ +360(1+a)^2\sqrt{5\theta}^2+60(1+a)^3\sqrt{5\theta}^3\end{array}\right]\\ +(b-a)^2\sqrt{5\theta}^2\left[150+120(1+a)\sqrt{5\theta}+30(1+a)^2\sqrt{5\theta}^2\right]\end{array}\right\}E_+\\
+\left\{\begin{array}{l}\left[\begin{array}{c}945+1350(1-a)\sqrt{5\theta}+810(1-a)^2\sqrt{5\theta}^2\\ +240(1-a)^3\sqrt{5\theta}^3+30(1-a)^4\sqrt{5\theta}^4\end{array}\right]\\ -(b-a)\sqrt{5\theta}\left[\begin{array}{c}675+810(1-a)\sqrt{5\theta}\\ +360(1-a)^2\sqrt{5\theta}^2+60(1-a)^3\sqrt{5\theta}^3\end{array}\right]\\ +(b-a)^2\sqrt{5\theta}^2\left[150+120(1-a)\sqrt{5\theta}+30(1-a)^2\sqrt{5\theta}^2\right]\end{array}\right\}E_-
\end{array}\right).$$

Additional factoring gives

$$J_{5/2}=\frac{1}{1080\sqrt{5\theta}}\left(\begin{array}{c}
2\left\{\begin{array}{c}945+945(b-a)\sqrt{5\theta}+420(b-a)^2\sqrt{5\theta}^2\\ +105(b-a)^3\sqrt{5\theta}^3+15(b-a)^4\sqrt{5\theta}^4+(b-a)^5\sqrt{5\theta}^5\end{array}\right\}E_\delta\\
-15\left\{\begin{array}{l}\left[\begin{array}{c}63+90(1+a)\sqrt{5\theta}+54(1+a)^2\sqrt{5\theta}^2\\ +16(1+a)^3\sqrt{5\theta}^3+2(1+a)^4\sqrt{5\theta}^4\end{array}\right]\\ +(b-a)\sqrt{5\theta}\left[\begin{array}{c}45+54(1+a)\sqrt{5\theta}+24(1+a)^2\sqrt{5\theta}^2\\ +4(1+a)^3\sqrt{5\theta}^3\end{array}\right]\\ +(b-a)^2\sqrt{5\theta}^2\left[10+8(1+a)\sqrt{5\theta}+2(1+a)^2\sqrt{5\theta}^2\right]\end{array}\right\}E_+\\
+15\left\{\begin{array}{l}\left[\begin{array}{c}63+90(1-a)\sqrt{5\theta}+54(1-a)^2\sqrt{5\theta}^2\\ +16(1-a)^3\sqrt{5\theta}^3+2(1-a)^4\sqrt{5\theta}^4\end{array}\right]\\ -(b-a)\sqrt{5\theta}\left[\begin{array}{c}45+54(1-a)\sqrt{5\theta}+24(1-a)^2\sqrt{5\theta}^2\\ +4(1-a)^3\sqrt{5\theta}^3\end{array}\right]\\ +(b-a)^2\sqrt{5\theta}^2\left[10+8(1-a)\sqrt{5\theta}+2(1-a)^2\sqrt{5\theta}^2\right]\end{array}\right\}E_-
\end{array}\right).$$

Collecting common powers of $\sqrt{5\theta}$ gives

$$J_{5/2} = \frac{1}{1080\sqrt{5\theta}} \left( \begin{array}{l} 2\left\{ \begin{array}{l} 945 + 945(b-a)\sqrt{5\theta} + 420(b-a)^2\sqrt{5\theta}^2 \\ +105(b-a)^3\sqrt{5\theta}^3 + 15(b-a)^4\sqrt{5\theta}^4 + (b-a)^5\sqrt{5\theta}^5 \end{array} \right\} E_\delta \\ -15\left\{ \begin{array}{l} 63 \\ +[90(1+a) + 45(b-a)]\sqrt{5\theta} \\ +[54(1+a)^2 + 54(b-a)(1+a) + 10(b-a)^2]\sqrt{5\theta}^2 \\ +\left[ \begin{array}{l} 16(1+a)^3 + 24(b-a)(1+a)^2 \\ +8(b-a)^2(1+a) \end{array} \right]\sqrt{5\theta}^3 \\ +[\ 2(1+a)^4 + 4(b-a)(1+a)^3 + 2(b-a)^2(1+a)^2]\sqrt{5\theta}^4 \end{array} \right\} E_+ \\ +15\left\{ \begin{array}{l} 63 \\ +[90(1-a) - 45(b-a)]\sqrt{5\theta} \\ +[54(1-a)^2 - 54(b-a)(1-a) + 10(b-a)^2]\sqrt{5\theta}^2 \\ +\left[ \begin{array}{l} 16(1-a)^3 - 24(b-a)(1-a)^2 \\ +8(b-a)^2(1+a) \end{array} \right]\sqrt{5\theta}^3 \\ +[\ 2(1-a)^4 - 4(b-a)(1-a)^3 + 2(b-a)^2(1-a)^2]\sqrt{5\theta}^4 \end{array} \right\} E_- \end{array} \right).$$

Factoring the square brackets further gives

$$J_{5/2} = \frac{1}{1080\sqrt{5\theta}} \left( \begin{array}{l} 2\left\{ \begin{array}{l} 945 + 945(b-a)\sqrt{5\theta} + 420(b-a)^2(5\theta) \\ +105(b-a)^3\sqrt{5\theta}^3 + 15(b-a)^4\sqrt{5\theta}^4 + (b-a)^5\sqrt{5\theta}^5 \end{array} \right\} E_\delta \\ -15\left\{ \begin{array}{l} 63 \\ +45[2(1+a) + (b-a)]\sqrt{5\theta} \\ +\ 2[27(1+a)^2 + 27(b-a)(1+a) + 5(b-a)^2]\sqrt{5\theta}^2 \\ +\ 8(1+a)[2(1+a)^2 + 3(b-a)(1+a) + (b-a)^2]\sqrt{5\theta}^3 \\ +\ 2(1+a)^2[(1+a)^2 + 2(b-a)(1+a) + (b-a)^2]\sqrt{5\theta}^4 \end{array} \right\} E_+ \\ +15\left\{ \begin{array}{l} 63 \\ +45[2(1-a) - (b-a)]\sqrt{5\theta} \\ +\ 2[27(1-a)^2 - 27(b-a)(1-a) + 5(b-a)^2]\sqrt{5\theta}^2 \\ +\ 8(1-a)[2(1-a)^2 - 3(b-a)(1-a) + (b-a)^2]\sqrt{5\theta}^3 \\ +\ 2(1-a)^2[(1-a)^2 - 2(b-a)(1-a) + (b-a)^2]\sqrt{5\theta}^4 \end{array} \right\} E_- \end{array} \right). \qquad (9.4)$$

---------- *Algebraic expansions for the penultimate and ultimate curly brackets of Eq. 9.4* ----------

$2(1 \pm a) \pm (b-a) = 2 \pm a \pm b.$

$$\begin{aligned} &27(1 \pm a)^2 \pm 27(b-a)(1 \pm a) + 5(b-a)^2 \\ &\quad = 27 \pm 54a + 27a^2 \pm 27b + 27ab \mp 27a \mp 27a^2 + 5b^2 - 10ab + 5a^2 \\ &\quad = 27 \pm 27a \pm 27b + 5a^2 + 5b^2 + 17ab. \end{aligned}$$

$$\begin{aligned} &(1 \pm a)[2(1 \pm a)^2 \pm 3(b-a)(1 \pm a) + (b-a)^2] \\ &\quad = (1 \pm a)(2 \pm 4a \pm 2a^2 \pm 3b + 3ab \mp 3a - 3a^2 + b^2 - 2ab + a^2) \\ &\quad = (1 \pm a)(2 \pm a \pm 3b + b^2 + ab) \\ &\quad = 2 \pm a \pm 3b + b^2 + ab \pm 2a + a^2 + 3ab \pm ab^2 \pm a^2 b \\ &\quad = 2 \pm 3a \pm 3b + b^2 + 4ab + a^2 \pm a^2 b \pm ab^2. \end{aligned}$$

$(1 \pm a)^2[(1 \pm a)^2 \pm 2(b-a)(1 \pm a) + (b-a)^2]$

$$
\begin{aligned}
&= (1 \pm a)^2[1 \pm 2a + a^2 \pm 2b \pm 2ab \mp 2a - 2a^2 + b^2 - 2ab + a^2] \\
&= (1 \pm a)^2[1 \pm 2b + b^2] \\
&= 1 \pm 2b + b^2 \pm 2a + 4ab \pm 2ab^2 + a^2 \pm 2a^2b + a^2b^2 \\
&= 1 \pm 2a \pm 2b + a^2 + 4ab + b^2 \pm 2a^2b \pm 2ab^2 + a^2b^2.
\end{aligned}
$$

---------- *End* ----------

Using the expansions in the above digression and substituting back in the expressions for $E_\delta$ and $E_\pm$ in Eqs. 9.2 and 9.3 gives

$$
J_{5/2} = \frac{1}{1080\sqrt{5\theta}} \left\{ \begin{array}{l}
2\left[\begin{array}{ll} 945 & \\ +945(b-a) & \sqrt{5\theta} \\ +420(b-a)^2 & \sqrt{5\theta}^2 \\ +105(b-a)^3 & \sqrt{5\theta}^3 \\ +\ \ 15(b-a)^4 & \sqrt{5\theta}^4 \\ +\qquad (b-a)^5 & \sqrt{5\theta}^5 \end{array}\right] e^{-2\sqrt{5\theta}\left|\frac{x_i-x_j}{2}\right|} \\
-15\left[\begin{array}{lll} 63 & & \\ +45 & \begin{pmatrix} 2 \\ +a+b \end{pmatrix} & \sqrt{5\theta} \\ +\ 2 & \begin{pmatrix} 27 \\ +27a+27b \\ +5a^2+5b^2+17ab \end{pmatrix} & \sqrt{5\theta}^2 \\ +\ 8 & \begin{pmatrix} 2 \\ +3a+3bv \\ +b^2+4ab+a^2 \\ +a^2b+ab^2 \end{pmatrix} & \sqrt{5\theta}^3 \\ +\ 2 & \begin{pmatrix} 1 \\ +2a+2b \\ +a^2+4ab+b^2 \\ +2a^2b+2ab^2 \\ +a^2b^2 \end{pmatrix} & \sqrt{5\theta}^4 \end{array}\right] e^{-2\sqrt{5\theta}\left(1+\frac{x_i+x_j}{2}\right)} \\
+15\left[\begin{array}{lll} 63 & & \\ +45 & \begin{pmatrix} 2 \\ -a-b \end{pmatrix} & \sqrt{5\theta} \\ +\ 2 & \begin{pmatrix} 27 \\ -27a-27b \\ +5a^2+5b^2+17ab \end{pmatrix} & \sqrt{5\theta}^2 \\ +\ 8 & \begin{pmatrix} 2 \\ -3a-3b \\ +b^2+4ab+a^2 \\ -a^2b-ab^2 \end{pmatrix} & \sqrt{5\theta}^3 \\ +\ 2 & \begin{pmatrix} 1 \\ -2a-2b \\ +a^2+4ab+b^2 \\ -2a^2b-2ab^2 \\ +a^2b^2 \end{pmatrix} & \sqrt{5\theta}^4 \end{array}\right] e^{-2\sqrt{5\theta}\left(1-\frac{x_i+x_j}{2}\right)}
\end{array} \right\}. \qquad (9.5)
$$

This agrees, as it should, with the result for this integral in Papers Ia, Sec. 5 of this series of papers [2], viz.,

$$
R^{(2)}_{i,j} = \frac{1}{1080\sqrt{5\theta_k}} \left\{ \begin{array}{l}
2\begin{bmatrix}
945 \\
+945\left|x_{i,k}-x_{j,k}\right| \sqrt{5\theta_k} \\
+420\left|x_{i,k}-x_{j,k}\right|^2 \sqrt{5\theta_k}^{2} \\
+105\left|x_{i,k}-x_{j,k}\right|^3 \sqrt{5\theta_k}^{3} \\
+\ 15\left|x_{i,k}-x_{j,k}\right|^4 \sqrt{5\theta_k}^{4} \\
+\quad \left|x_{i,k}-x_{j,k}\right|^5 \sqrt{5\theta_k}^{5}
\end{bmatrix} e^{-2\sqrt{5\theta_k}\left|\frac{x_{i,k}-x_{j,k}}{2}\right|} \\
-15\mathcal{T}^{(-)}_{x_{i,k};x_{j,k}} \begin{bmatrix}
63 \\
+45\begin{pmatrix} 2 \\ +x_{i,k}+x_{j,k} \end{pmatrix} \sqrt{5\theta_k} \\
+\ 2\begin{pmatrix} 27 \\ +27x_{i,k}+27x_{j,k} \\ +\ 5x_{i,k}^2+17x_{i,k}x_{j,k}+5x_{j,k}^2 \end{pmatrix} \sqrt{5\theta_k}^{2} \\
+\ 8\begin{pmatrix} 2 \\ +3x_{i,k}+3x_{j,k} \\ +x_{i,k}^2+4x_{i,k}x_{j,k}+x_{j,k}^2 \\ +x_{i,k}^2x_{j,k}+x_{i,k}x_{j,k}^2 \end{pmatrix} \sqrt{5\theta_k}^{3} \\
+\ 2\begin{pmatrix} 1 \\ +2x_{i,k}+2x_{j,k} \\ +x_{i,k}^2+4x_{i,k}x_{j,k}+x_{j,k}^2 \\ +2x_{i,k}^2x_{j,k}+2x_{i,k}x_{j,k}^2 \\ +x_{i,k}^2x_{j,k}^2 \end{pmatrix} \sqrt{5\theta_k}^{4}
\end{bmatrix} e^{-2\sqrt{5\theta_k}\left(1+\frac{x_{i,k}+x_{j,k}}{2}\right)}
\end{array} \right\}.
$$

## 10. Bessel numbers of the first kind

We conjecture, and leave the proof to the interested researcher, that the integer coefficients in the top curly brackets of Eqs. 8.14 and 8.15 are signless Bessel numbers of the first kind [12,13]. Some of the relevant Bessel numbers of the first kind are tabulated in Table 3, immediately below.

| ν | $a_0$ | $a_1$ | $a_2$ | $a_3$ | $a_4$ | $a_5$ | $a_6$ | $a_7$ | $a_8$ | $a_9$ |
|---|---|---|---|---|---|---|---|---|---|---|
| **3/2** | **1** | **6** | **15** | **15** | | | | | | |
| 2 | 1 | 10 | 45 | 105 | 105 | | | | | |
| **5/2** | **1** | **15** | **105** | **420** | **945** | **945** | | | | |
| 3 | 1 | 21 | 210 | 1260 | 4725 | 10395 | 10395 | | | |
| **7/2** | 1 | 28 | 378 | 3150 | 17325 | 62370 | 135135 | 135135 | | |
| 4 | 1 | 36 | 630 | 6930 | 51975 | 270270 | 945945 | 2027025 | 2027025 | |
| **9/2** | 1 | 45 | 990 | 13860 | 135135 | 945945 | 4729725 | 16216200 | 34459425 | 34459425 |

Table 3. Bessel numbers of the first kind, from the OEIS [11], are tabulated for $3/2 \le \nu \le 9/2$. The entries in the row with heading $\nu = 3/2$, viz., $[1, 6, 15, 15]$, are the integer coefficients in the top square bracket of Eq. 9.1; while the entries in the row with heading $\nu = 5/2$, viz., $[1, 15, 105, 420, 945, 945]$, are the integer coefficients in the top square bracket of Eq. 9.5.

## 11. Summary and concluding comments

Eq. 6.6 (respectively, Eqs. 8.14 and 8.15) provides hand-generated algebraic expressions of integrals of single (resp., products of two) Matérn-covariance functions, for all odd-half-integer class parameters. Some derived coefficients in these integrals are conjectured to be Bessel numbers of the first kind.

## 12. Research reproducibility

We support the recommendations of ICERM's Workshop on Reproducibility in Computational and Experimental Mathematics Workshop [14]. All data and figure-generation files used in this research are available to responsible parties from the first author at selden_crary (at) yahoo (dot) com.

## 13. Version history

V2: "Generalized" was added to the title. The numeration of tables was corrected.

## Acknowledgments

We thank Arun Jambulapati of Stanford Univ.'s Institute for Computational & Mathematical Engineering for pointing us to [5]. We also thank Prof. Max Morris of Iowa State Univ. for supportive comments.

## Appendix: A proof of the Variation of Finite Companion Binomial theorem

We now provide a proof of the variation of finite companion binomial theorem of Sec. 5.2, in an experimental, highly graphical form. The first author takes full responsibility for any abuse of notation.

**Statement of the theorem**

To show: $\sum_{j=0}^{p} \frac{(2p-j)!}{(p-j)!} 2^j = 4^p p!\,, \;\; p \in \mathbb{N}_{\geq 0}$.

**Preliminaries**

Rows 0 through 11 of the unmodified Pascal's triangle (hereafter 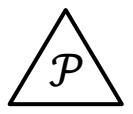) are shown below, along with the numbering scheme for rows, diagonals, and anti-diagonals:

```
        Row 0 -->          1
        Row 1 -->        1   1
        Row 2 -->      1   2   1
                     1   3   3   1
                   1   4   6   4   1
                 1   5  10  10   5   1
               1   6  15  20  15   6   1
             1   7  21  35  35  21   7   1
           1   8  28  56  70  56  28   8   1
         1   9  36  84 126 126  84  36   9   1
       1  10  45 120 210 252 210 120  45  10   1
     1  11  55 165 330 462 462 330 165  55  11   1

Anti-diag. 0 -->  Anti-diag. 1 -->  Anti-diag. 2 -->  ...
...  <-- Diagonal 2  <-- Diagonal 1  <-- Diagonal 0
```

**Demonstration**

Starting from the binomial coefficients of the Binomial theorem, $\binom{a}{b} \equiv \frac{a!}{b!(a-b)!}$ , $a \geq b$,

$$\sum_{j=0}^{p} \frac{(2p-j)!}{(p-j)!} 2^j = p! \sum_{j=0}^{p} 2^j \binom{2p-j}{p} = p! \left[ \binom{2p}{p} + 2\binom{2p-1}{p} + \sum_{j=2}^{p} 2^j \binom{2p-j}{p} \right] \tag{1}$$

$$= p! \left[ \frac{2p(2p-1)!}{p!p!} + 2\frac{(2p-1)!}{(p-1)!p!} + \sum_{j=2}^{p} 2^j \binom{2p-j}{p} \right]$$

$$= p! \left[ \frac{2p(2p-1)!}{p!p!} + \frac{2p(2p-1)!}{p!p!} + \sum_{j=2}^{p} 2^j \binom{2p-j}{p} \right]$$

$$= p! \left[ \frac{4(2p-1)!}{(p-1)!p!} + \sum_{j=2}^{p} 2^j \binom{2p-j}{p} \right] = p! \left[ 4\binom{2p-1}{p} + \sum_{j=2}^{p} 2^j \binom{2p-j}{p} \right]$$

$$= 4 \cdot p! \left[ \binom{2p-1}{p} + \sum_{j=2}^{p} 2^{j-2} \binom{2p-j}{p} \right],$$

so it remains to show $\binom{2p-1}{p} + \sum_{j=2}^{p} 2^{j-2} \binom{2p-j}{p} = 4^{p-1}$. (2)

The $p$ binomial coefficients of Eq. 2 lie along Anti-diagonal $p$ in Rows $p$ through $2p-1$ of 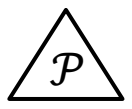, with the modification that the Row $p$ coefficient is multiplied by $2^{p-2}$, and the relevant coefficients in successive rows are multiplied by $2^{p-3}, 2^{p-4}, \cdots, 2, 2^0$, and unity (again). To assist the student, we carry forward the needed demonstration by writing out the relevant rows of 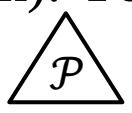 for the $p = 6$ case, with red used for the (modified) coefficients in Eq. 1. We seek to show the sum of the red cells in the triangular tableau is $4^{p-1}$, i.e. that it is equal to the sum of the unmodified binomial coefficients of Row $(2p-2)$. That is, we seek to show the sum of red integers in 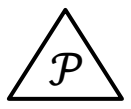 segments "a", below, equals the sum of green integers in 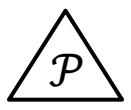 segments "b", below.

```
(a)        1   6  15  20  15   6  16
         1   7  21  35  35  21  56   1
       1   8  28  56  70  56 112   8   1
     1   9  36  84 126 126 168  36   9   1
   1  10  45 120 210 252 210 120  45  10   1
 1  11  55 165 330 462 462 330 165  55  11   1
```

```
(b)        1   6  15  20  15   6   1
         1   7  21  35  35  21   7   1
       1   8  28  56  70  56  28   8   1
     1   9  36  84 126 126  84  36   9   1
   1  10  45 120 210 252 210 120  45  10   1
 1  11  55 165 330 462 462 330 165  55  11   1
```

Graphic above. For the case $p = 6$, Rows 6 through 11 of $\triangle_{\mathcal{P}}$, are relevant and are shown in each of (a) and (b), above. In (a), the black integers are $\triangle_{\mathcal{P}}$ entries, while the red integers are $\triangle_{\mathcal{P}}$ entries multiplied by $2^4, 2^3, \cdots, 2^0$, or unity; for Rows $6, 7, \cdots, 11$; respectively. In (b), the integers are $\triangle_{\mathcal{P}}$ ,with green used for the Row 10 entries. For the case at hand, the identity sought in this paper is the equality of the sum of the colored entries in each of (a) and (b) to $4^{p-1} = 4^5 = 1024$.

It will be useful to show the modifications of the entries of $\triangle_{\mathcal{P}}$ explicitly, as in the following:

```
               1     6    15    20    15     6  2⁴·1
            1     7    21    35    35    21  2³·7     1
         1     8    28    56    70    56  2²·28     8     1
      1     9    36    84   126   126  2¹·84    36     9     1
   1    10    45   120   210   252  2⁰·210   120    45    10     1
1    11    55   165   330   462   1·462   330   165    55    11     1.
```

The coefficient of interest in Row $2p-1$ (the bold 462 in the truncated triangle, above) is the latter of the two equal central coefficients in that row, viz. $\binom{2p-1}{p-1}$ and $\binom{2p-1}{p}$. By the defining recursion of $\triangle_{\mathcal{P}}$, the former is equal to $\binom{2p-2}{p-2} + \binom{2p-2}{p-1}$, so we can, in our sum of sometimes modified coefficients, replace the coefficient $\binom{2p-1}{p-1}$ with two others, viz. $\binom{2p-2}{p-2}$ and $\binom{2p-2}{p-1}$. Then, removing the last row, as it will not be needed subsequently, the colored, truncated $\triangle_{\mathcal{P}}$ becomes

```
               1     6    15    20    15     6  2⁴·1
            1     7    21    35    35    21  2³·7     1
         1     8    28    56    70    56  2²·28     8     1
      1     9    36    84   126   126  2¹·84    36     9     1
   1    10    45   120   210   252   210   120    45    10     1
```

The values of the upper $p-2$ modified coefficients along the $p$'th anti-diagonal can be split evenly over their own cell in $\overset{\mathcal{P}}{\triangle}$ and the mirror-image location, across a vertical bisector of $\overset{\mathcal{P}}{\triangle}$, giving, using an obvious definition of $Cell\binom{i}{j}$ as the contents of the Row $i$ and Anti-diagonal $j$ bisector of $\overset{\mathcal{P}}{\triangle}$ :

$$Cell\binom{p}{p} = Cell\binom{p}{0} = \frac{1}{2}2^{p-2}\binom{p}{p} = 2^{p-3},$$

$$Cell\binom{p+1}{p} = Cell\binom{p+1}{1} = \frac{1}{2}2^{p-3}\binom{p+1}{p} = 2^{p-4}\binom{p+1}{p},$$

$$\vdots$$

$$Cell\binom{2p-3}{p} = Cell\binom{p-1}{p-3} = \frac{1}{2}2^{1}\binom{2p-3}{p} = 2^{0}\binom{2p-3}{p}.$$

This leads to the graphic,

$$
\begin{array}{ccccccccccccccccccccc}
 & & & & 2^3\cdot\mathbf{1} & & 6 & & 15 & & 20 & & 15 & & 6 & & 2^3\cdot\mathbf{1} & & & & \\
 & & & 1 & & 2^2\cdot\mathbf{7} & & 21 & & 35 & & 35 & & 21 & & 2^2\cdot\mathbf{7} & & 1 & & & \\
 & & 1 & & 8 & & 2^1\cdot\mathbf{28} & & 56 & & 70 & & 56 & & 2^1\cdot\mathbf{28} & & 8 & & 1 & & \\
 & 1 & & 9 & & 36 & & 2^0\cdot\mathbf{84} & & 126 & & 126 & & 2^0\cdot\mathbf{84} & & 36 & & 9 & & 1 & \\
1 & & 10 & & 45 & & 120 & & \mathbf{210} & & \mathbf{252} & & \mathbf{210} & & 120 & & 45 & & 10 & & 1.
\end{array}
$$

At this stage, the $p$'th row consists of the single entry in the uppermost left location and an equal entry in the uppermost right location, with the latter being on Anti-diagonal $p$ and having value $2^{p-3}\binom{p}{p}$. Successive entries, descending along the anti-diagonal, are $2^{p-4}\binom{p+1}{p}, 2^{p-5}\binom{p+2}{p}, \cdots, 2^{0}\binom{2p-3}{p}$, and $\binom{2p-2}{p}$.

Now familiar with the steps that can be applied to the graphic, including the defining recursion of $\overset{\mathcal{P}}{\triangle}$, the student can follow the steps below that lead to the desired result:

$$
\begin{array}{ccccccccccccccccccccc}
 & & & & & 1 & & 6 & & 15 & & 20 & & 15 & & 6 & & 1 & & & \\
 & & & 2^3\cdot\mathbf{1} & & 2^2\cdot\mathbf{7} & & 21 & & 35 & & 35 & & 21 & & 2^2\cdot\mathbf{7} & & 2^3\cdot\mathbf{1} & & & \\
 & & 1 & & 8 & & 2^1\cdot\mathbf{28} & & 56 & & 70 & & 56 & & 2^1\cdot\mathbf{28} & & 8 & & 1 & & \\
 & 1 & & 9 & & 36 & & 2^0\cdot\mathbf{84} & & 126 & & 126 & & 2^0\cdot\mathbf{84} & & 36 & & 9 & & 1 & \\
1 & & 10 & & 45 & & 120 & & \mathbf{210} & & \mathbf{252} & & \mathbf{210} & & 120 & & 45 & & 10 & & 1,
\end{array}
$$

1 6 15 20 15 6 1

1 7 21 35 35 21 7 1

1 8 $2^1 \cdot 28$ 56 70 56 $2^1 \cdot 28$ $2^2 \cdot 8$ $2^2 \cdot 1$

1 9 36 $2^0 \cdot 84$ 126 126 $2^0 \cdot 84$ 36 9 1

1 10 45 120 **210** **252** **210** 120 45 10 1,

1 6 15 20 15 6 1

1 7 21 35 35 21 7 1

1 8 28 56 70 56 28 8 1

$2^1 \cdot 1$ $2^1 \cdot 9$ 36 $2^0 \cdot 84$ 126 126 $2^0 \cdot 84$ $2^1 \cdot 36$ $2^1 \cdot 9$ $2^1 \cdot 1$

1 10 45 120 **210** **252** **210** 120 45 10 1,

and finally,

1 6 15 20 15 6 1

1 7 21 35 35 21 7 1

1 8 28 56 70 56 28 8 1

1 9 36 84 126 126 84 36 9 1

**1 10 45 120 210 252 210 120 45 10 1**,

which was to be shown.

The astute student will see that this graphical procedure carries through, in general.